\documentclass{aastex61}

\newcommand\aastex{AAS\TeX}

\renewcommand{\thefootnote}{\fnsymbol{footnote}}

\shorttitle{\aastex\ sample article}
\shortauthors{Devillepoix et al.}

\begin{document}

\title{The Dingle Dell meteorite: a Halloween treat from the Main Belt}

\correspondingauthor{Hadrien A. R. Devillepoix}
\affiliation{School of Earth and Planetary Sciences, Curtin University, Bentley, WA 6102, Australia}
\email{h.devillepoix@postgrad.curtin.edu.au}

\author[0000-0001-9226-1870]{Hadrien A. R. Devillepoix}
\affiliation{School of Earth and Planetary Sciences, Curtin University, Bentley, WA 6102, Australia}

\author[0000-0003-2702-673X]{Eleanor K. Sansom}
\affiliation{School of Earth and Planetary Sciences, Curtin University, Bentley, WA 6102, Australia}

\author{Philip A. Bland}
\affiliation{School of Earth and Planetary Sciences, Curtin University, Bentley, WA 6102, Australia}

\author[0000-0002-8240-4150]{Martin C. Towner}
\affiliation{School of Earth and Planetary Sciences, Curtin University, Bentley, WA 6102, Australia}

\author{Martin Cup\'{a}k}
\affiliation{School of Earth and Planetary Sciences, Curtin University, Bentley, WA 6102, Australia}

\author[0000-0002-5864-105X]{Robert M. Howie}
\affiliation{School of Earth and Planetary Sciences, Curtin University, Bentley, WA 6102, Australia}

\author[0000-0002-0363-0927]{Trent Jansen-Sturgeon}
\affiliation{School of Earth and Planetary Sciences, Curtin University, Bentley, WA 6102, Australia}

\author[0000-0001-9308-0123]{Morgan A. Cox}
\affiliation{School of Earth and Planetary Sciences, Curtin University, Bentley, WA 6102, Australia}

\author[0000-0002-8646-0635]{Benjamin A. D. Hartig}
\affiliation{School of Earth and Planetary Sciences, Curtin University, Bentley, WA 6102, Australia}

\author[0000-0003-0990-8878]{Gretchen K. Benedix}
\affiliation{School of Earth and Planetary Sciences, Curtin University, Bentley, WA 6102, Australia}

\author{Jonathan P. Paxman}
\affiliation{School of Civil and Mechanical Engineering, Curtin University, Bentley, WA 6102, Australia}

\begin{abstract}

We describe the fall of the Dingle Dell (L/LL\,5) meteorite near Morawa in Western Australia on October 31, 2016.
The fireball was observed by six observatories of the Desert Fireball Network (DFN), a continental scale facility optimised to recover meteorites and calculate their pre-entry orbits.
The $30\,\mbox{cm}$ meteoroid entered at 15.44\,$\mbox{km s}^{-1}$, followed a moderately steep trajectory of $51\degr$ to the horizon from 81\,km down to 19\,km altitude, where the luminous flight ended at a speed of 3.2\,$\mbox{km s}^{-1}$.
Deceleration data indicated one large fragment had made it to the ground.
The four person search team recovered a 1.15\,kg meteorite within 130\,m of the predicted fall line, after 8 hours of searching, 6 days after the fall.
Dingle Dell is the fourth meteorite recovered by the DFN in Australia, but the first before any rain had contaminated the sample.
By numerical integration over 1 Ma, we show that Dingle Dell was most likely ejected from the main belt by the 3:1 mean-motion resonance with Jupiter, with only a marginal chance that it came from the $nu_6$ resonance. This makes the connection of Dingle Dell to the Flora family (currently thought to be the origin of LL chondrites) unlikely.

\end{abstract}


\section{Introduction} \label{sec:intro}
As of mid-2017 there are nearly 60k meteorite samples classified in the Meteoritical Bulletin Database\footnote{\url{https://www.lpi.usra.edu/meteor/metbull.php}}.
However, apart from a handful of Lunar ($\simeq300$) and Martian ($\simeq200$) meteorites that have a well known origin, the link with other solar system bodies is limited.
From the instrumentally documented fall of the P\v{r}\'{\i}bram meteorite in 1959 \citep{1961BAICz..12...21C}, we learned that chondritic material comes from the asteroid main belt.
The way this material evolves onto an Earth crossing orbit starts with a disruption in the main belt. 
The small members of the debris field can be strongly affected by the Yarkovsky effect \citep{1998Icar..132..378F} and as a consequence their semi-major axis is continually altered.
If the debris field is close to a powerful resonance (in semi-major axis, inclination, eccentricity space), the break up event feeds material into that resonance, which will in turn push the debris' perihelia into the inner solar system. This can occur on a timescale of less than a million years in some cases \citep{1994A&A...282..955M}.

Calculating the orbit of a meteoroid using only the luminous trajectory as the observation arc is in most cases not precise enough to allow unequivocal backtracking into a specific region of the main belt, hence the statistical results reported by  \cite{2009Sci...325.1525B,2011M&PS...46..339B,2014M&PS...49.1388J,2015MNRAS.449.2119T}. In order to understand the origin of the different groups of meteorites from the main asteroid belt, it is therefore essential to collect several dozen samples with orbits and look at source regions in a broader, statistical way.

\subsection{Dedicated networks to recover meteorites with known provenance}

In the decade following 2000, the recovery rate of meteorites with determined orbits has dramatically increased \citep{2015aste.book..257B}, without a significant increase in collecting area of the major dedicated fireball networks.
While the initial phase of the Desert Fireball Network (DFN) started science operations in December 2005, covering $0.2\times10^{6}\,\mbox{km}^2$ \citep{2012AuJES..59..177B},
other major networks ceased operations.
The Prairie network in the USA ($0.75\times10^{6}\,\mbox{km}^2$ \citep{McCrosky1965}) shut down in 1975,
the Canadian Meteorite Observation and Recovery Project (MORP) - $1.3\times10^{6}$\,km$^2$- stopped observing in 1985 \citep{1996M&PS...31..185H},
and the European Network's covering area of $\sim1\times10^{6}\,\mbox{km}^2$ has not significantly changed \citep{1998M&PS...33...49O}.
If not due to a larger collecting area, this increase can be explained by other factors:
\begin{itemize}
    \item Existing networks improving their data reduction techniques \citep{2014A&A...570A..39S}.
    \item Democratisation and cheap operating cost of recording devices (surveillance cameras, consumer digital cameras...) \citep{2003M&PS...38..975B}.
    \item Use of doppler radar designed for weather observations to constrain the location of falling meteorites  \citep{2012Sci...338.1583J,2014M&PS...49.1989F,2010M&PS...45.1476F}.
    \item Deployment of the Desert Fireball Network expressly on favourable terrain to search for meteorites. In its early stage, within its first 5 years of science operation, the DFN yielded 2 meteorites \citep{2009Sci...325.1525B,2011M&PSA..74.5101S}, whilst MORP only yielded one \citep{1981Metic..16..153H} in 15 years of operations over a larger network.
    \item To a lesser extent, development of NEO telescopic surveillance programmes. One single case so far (the Catalina Sky Survey detecting the Almahata Sita meteoroid several hours before impact \citet{2009Natur.458..485J}), however this technique is likely to yield more frequent successes with new deeper and faster optical surveyors, like LSST, which comes online in 2021 \citep{2008SerAJ.176....1I}.
\end{itemize}

The DFN started developing digital observatories to replace the film based network in 2012 with the goal of covering $10^{6}\,\mbox{km}^2$, the more cost effective than expected digital observatories allowed the construction of a continent-scale network covering over $2.5\times10^{6}\,\mbox{km}^2$ \citep{2017ExA...tmp...19H}.
This programme rapidly yielded results, less than a year after starting science operation (in November 2014). One of the observatories lent to the SETI institute in California was a crucial viewpoint to calculating an orbit for the Creston fall in California in October 2015 \citep{Metbull_104}, and the first domestic success came 2 months later with the Murrili meteorite recovery on Kati Thanda--Lake Eyre \citep{2016pimo.conf...60D,Metbull_105}.
We report here the analysis of observations of a bright fireball that led to the fourth find by the Desert Fireball Network in Australia: the Dingle Dell meteorite.  Dingle Dell was originally classified as an LL ordinary chondrite, petrographic type 6 \citep{Metbull_106}. However, further analysis revealed that it in fact sits on the L/LL boundary \citep{benedix2017dd}. The sample has experienced a low level of shock, but has been heated enough to show recrystallisation of minerals and matrix. There is no evidence of terrestrial weathering visible on the metal or sulphide grains, which is consistent with its extremely fast retrieval from the elements.

\subsection{Current understanding of the origin of the main groups of L and LL chondrites}

\paragraph{L chondrites}

L chondrites represent 32\% of total falls.
\citet{2001E&PSL.194....1S} first identified a large amount of fossil L chondrites meteorites in $\simeq467\,\mbox{Ma}$ sedimentary rock, which suggests that a break up happened not too long before, near an efficient meteorite transport route.
From spectroscopic and dynamical arguments, \citet{2009Icar..200..698N} proposed that the Gefion family break up event, close to the 5:2 MMR with Jupiter, might be the source of this bombardment, given the rapid delivery time, and a likely origin of L chondrite asteroids outside of the 2.5 AU.
Most shocked L5 and L6 instrumentally observed falls also seem to come from this break up, with an $^{39}Ar - ^{40}Ar$ age around $\simeq470$\,Ma ago: Park Forest \citep{2004M&PS...39.1781B},
Novato \citep{2014M&PS...49.1388J},
Jesenice \citep{2010M&PS...45.1392S},
and Innisfree \citep{1981Metic..16..153H}.
Only the Villalbeto de la Pe{\~n}a L6 \citep{2006M&PS...41..505T} does not fit in this story because of its large cosmic ray exposure age (48 Ma), inconsistent with a 8.9 Ma collisional lifetime \citep{2014me13.conf...57J}.

\paragraph{LL chondrites}
Thanks to \citet{2008Natur.454..858V}, we know that S- and Q-type asteroids observed in NEO space are the most likely asteroidal analogue to LL type ordinary chondrites.
The Hayabusa probe returned samples from S-type (25143)\,Itokawa, finally unequivocally matching the largest group of meteorites recovered on Earth (ordinary chondrites) with the most common spectral class of asteroids in the main belt \citep{2011Sci...333.1113N}. 
The sample brought back from Itokawa is compatible with LL chondrites.
Indeed, LL compatible asteroids make up two thirds of near-Earth space. The spectrally compatible Flora family from the inner main belt can regenerate this population through the $\nu_6$ secular resonance.
But one large problem remains: only 8\% of falls are LL chondrites \citep{2008Natur.454..858V}.
The orbits determined for some LL samples have so far not helped solve this issue.
If we exclude Bene{\v s}ov \citep{2014A&A...570A..39S}, which was a mixed fall, scientists had to wait until 2013 to get an LL sample with a precisely calculated orbit: Chelyabinsk \citep{2013Natur.503..238B,2013Natur.503..235B}.
The pre-atmospheric orbit and composition of the Chelyabinsk meteorite seems to support the Flora family origin for LL chondrites, although a more recent impact could have reset the cosmic ray exposure age to $1.2\pm0.2$\,Ma, and the presence of impact melts (very rare in ordinary chondrites due to the large impact velocities required \citep{1997M&PS...32..349K}).
\citet{2014Icar..237..116R} argued that an impact melt such as the one observed in the Chelyabinsk meteorites, or shock darkening, can alter the spectra of an S/Q-type asteroid to make it look like a C/X-type spectrally.
The implication of this is that the Baptistina family members (C/X-type), which overlaps dynamically with the Flora (S-type), could be the remains of a large impact on a Florian asteroid, and meteorites from both families can be confused both in their spectral signature and dynamical origin.
It must be noted however that \citet{2014Icar..237..116R} do not make any conclusions on the origin of Chelyabinsk from the Baptistina family. The Chelyabinsk meteorite is also not a typical LL sample found on Earth, because of its size ($\simeq17\,m$), and the presence of impact melts.

Based on it's classification, we put the orbit of the Dingle Dell meteorite in context with other calculated orbits from L and LL chondrites and discuss the resonances from which it may have originated.

\section{Fireball observation and trajectory data}

\begin{figure}[!h]
    \centering
    \includegraphics[width=0.8\linewidth]{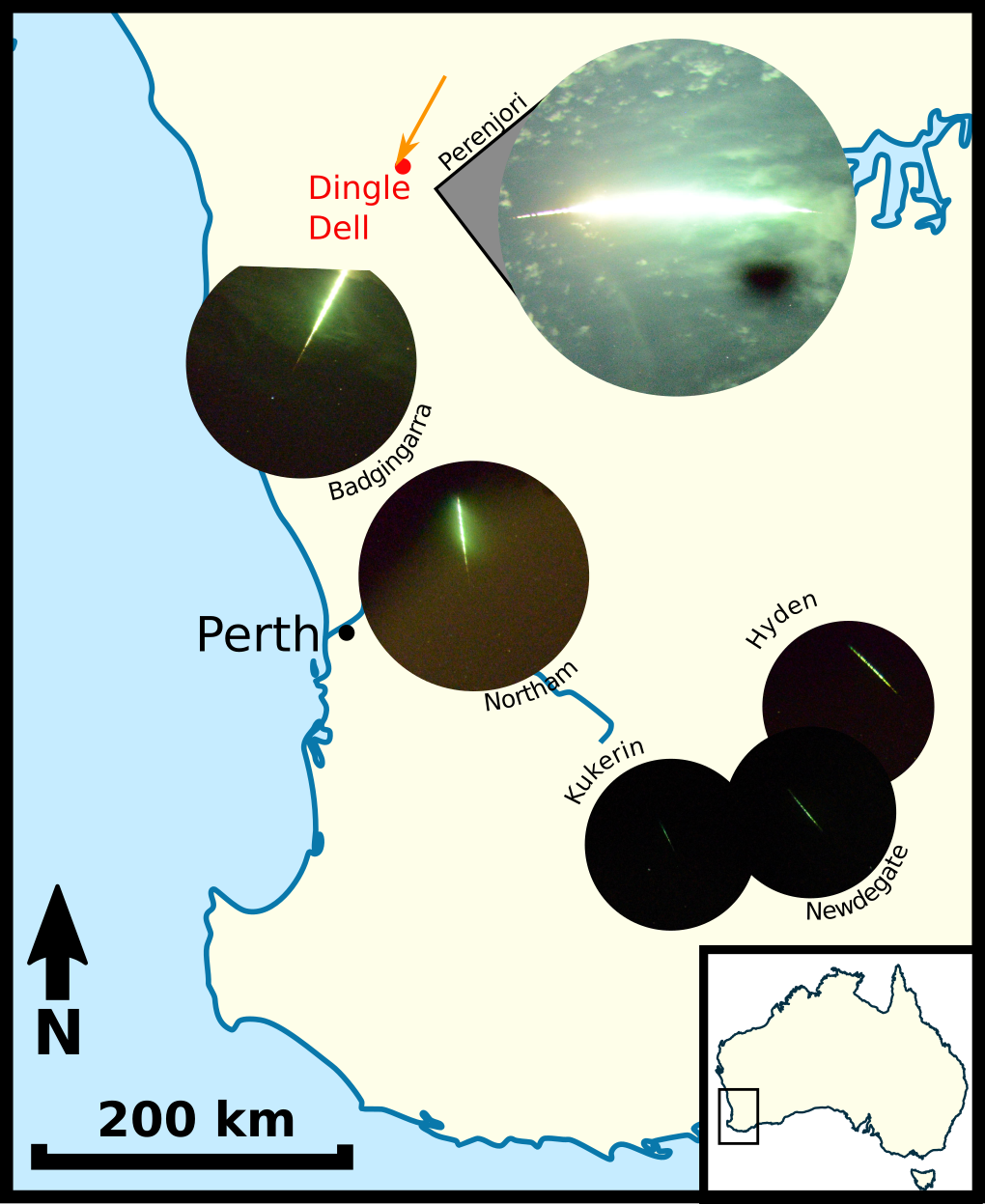}
    \caption{Cropped all-sky images of the fireball from the six DFN observatories. Images are of the same pixel scale with the centre of each image positioned at the observatory location on the map (with the exception of Perenjori, whose location is indicated).
    The Badgingarra image is cropped because the sensor is not large enough to accommodate the full image circle on its short side.
    The saturation issue is exacerbated by light scattered in the clouds on cameras close to the event, this is particularly visible on the Perenjori image.
    The black blotch in the Perenjori image is an artefact that thankfully did not extend far enough to affect the quality of the data. 
     Approximate trajectory path shown by orange arrow.
     Location of the recovered meteorite is shown by the red dot.}
    \label{fig:fireball}
\end{figure}

On Halloween night shortly after 8 PM local time, several reports of a large bolide were made via the \textit{Fireballs In The Sky} smart-phone app \citep{2016pimo.conf..267S} from the Western Australian Wheatbelt area.
These were received a few hours prior to the daily DFN observatory reports, apprising the team of the event expeditiously. 
The DFN observatory sightings are routinely emailed after event detection has been completed on the nights' data-set. It revealed that six nearby DFN observatories simultaneously imaged a long fireball starting at 12:03:47.726 UTC on October 31, 2016 (Figure \ref{fig:fireball}).

\subsection{Instrumental records}

The main imaging system of the DFN fireball observatories is a 36 MPixel sensor: Nikon D810 (or D800E on older models), combined with a Samyang lens 8mm F/3.5.
Long exposure images are taken every 30 seconds. The absolute and relative timing (from which the fireball velocity is derived) is embedded into the luminous trail by use of a liquid crystal (LC) shutter between the lens and the sensor, modulated according to a de-Brujin sequence \citep{2017M&PS...52.1669H}. The LC shutter operation is tightly regulated by a micro-controller synced with a Global Navigation Satellite System (GNSS) module to ensure absolute timing accurate to $\pm0.4\,\mbox{ms}$. For further details on DFN observatory specifications, see \citet{2017ExA...tmp...19H}. 

Some DFN observatories also include video systems operating in parallel with the long exposure photographic imaging system (Table \ref{table:stations}). The video cameras are Watec 902H2 Ultimate CCIR (8 bit 25 interlaced frames per second), with a Fujinon fisheye lens.
Originally intended as a backup device for absolute timing, these video systems have been retained for future daytime observation capabilities. Here we make use of the video data to acquire a light curve, as the event saturated the still camera sensors.
The closest camera system to this event was in Perenjori (Table \ref{table:stations}), located almost directly under the fireball, and was the only station to image the end of the luminous trajectory (Fig. \ref{fig:fireball}). Other nearby camera sites were
overcast and did not record the event. In order to triangulate the trajectory of the fireball, distant stations had to be used, all over 200 km away. The Hyden, Kukerin and Newdegate systems were all around 500 km from the event and, although still managing to capture the fireball, were too low on the horizon for accurate calibration.

\begin{table}[!h]
	\caption{Locations and nature of instrumental records. We use cameras $<400\,km$ away for trajectory determination.}              % title of Table
	\label{table:stations}      % is used to refer this table in the text
	\centering                                      % used for centering table
	\begin{tabular}{l c c c c c}          % centered columns (4 columns)
		\hline\hline                        % inserts double horizontal lines
		Observatory & Instruments & Latitude & Longitude & altitude (m) & distance \tablenotemark{*} (km) \\
		\hline      
		Perenjori & P, V &  29.36908 S & 116.40654 E & 242  & 91  \\ 
		Badgingarra  & P &  30.40259 S & 115.55077 E & 230  & 204  \\ 
		Northam  & P &  31.66738 S & 116.66571 E & 190  & 323  \\ 
		\hline
		Hyden  & P &  32.40655 S & 119.15325 E & 390  & 484  \\  
		Kukerin  & P &  33.25337 S & 118.00628 E & 340  & 520  \\  
		Newdegate  & P &  33.05436 S & 118.93534 E & 302  & 534  \\  
		\hline                                             %inserts single line
	\end{tabular}
		\tablenotetext{}{P: Photographic record (exposures: 25 seconds, 6400 ISO, F/4.), V: video record}
		\tablenotetext{*}{distance from the meteoroid at 70 km altitude}
	
\end{table}

\begin{figure}[h!]
    \centering
    \includegraphics[width=\linewidth]{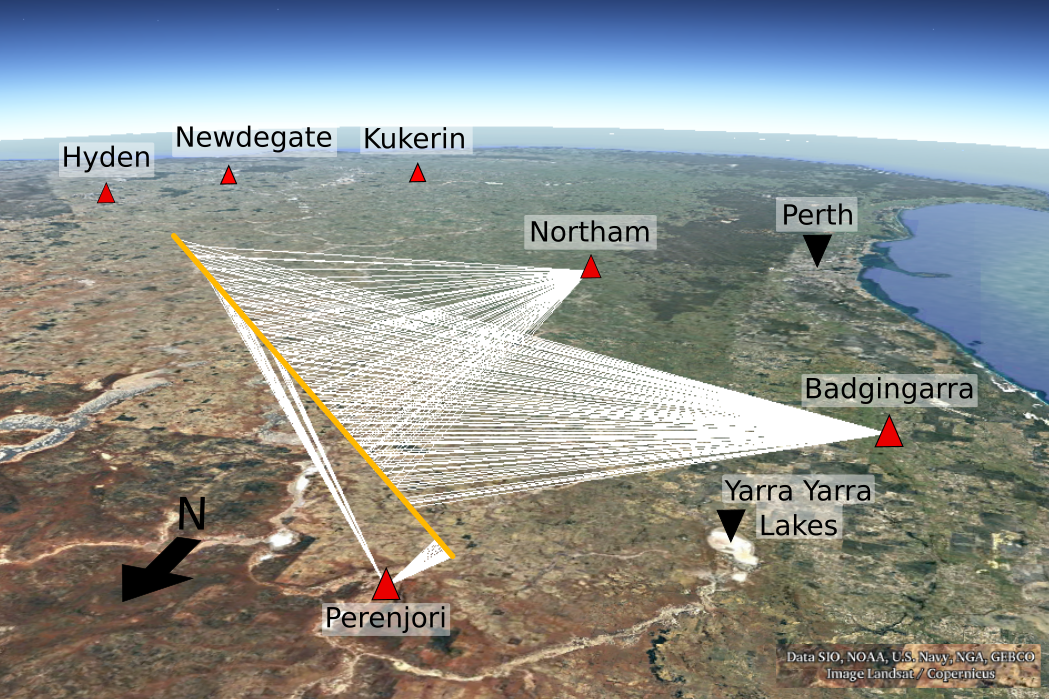}
    \caption{Configuration of DFN station observations for the Dingle Dell fireball. White rays show observations used in triangulation of the trajectory (approximated to the yellow line, starting NE and terminating to the SW of Perenjori). Hyden, Newdegate and Kukerin stations were all around 500 km away from the event and were not used in triangulation.}
    \label{fig:g_earth_rays}
\end{figure}

\subsection{Astrometry}\label{sec:astrometry}
All images captured by the DFN observatories are saved even when no fireball is detected.
This is possible thanks to the availability of large capacity hard drives at reasonable costs.
Not only does this mitigate event loss during initial testing of detection algorithms, but it 
gives a snapshot of the whole visible sky down to 7.5 point source limiting magnitude, every 30 seconds. The astrometric calibration allows the points picked along the fireball image to be converted to astrometric sky coordinates. The associated astrometric uncertainties are dominated by the uncertainty on identifying the centroids along the segmented fireball track.

We have carried out studies on the long-term camera stability by checking the camera pointing using astrometry.
On the outback system tested, the pointing changed less than $1\arcmin$ over the 3 month period assessed.
The pointing is therefore remarkably stable, and the relevant fireball image can thus be astrometrically calibrated using a picture taken at a different epoch. This is particularly useful when a bright fireball overprints nearby stars, and especially in this case where clouds are present.
In general however, we aim to use a calibration frame taken as close as possible from the science frame, particularly when studying an important event, such as a meteorite fall.
In the following paragraph we present the methods used for astrometrically calibrating the still images, using as an example the Perenjori data. This technique is implemented in an automated way in the reduction pipeline for all detected events.% and is detailed further in \citet{hadry2017taurids}.

\begin{figure}[!h]
    \centering
    \includegraphics[width=\linewidth]{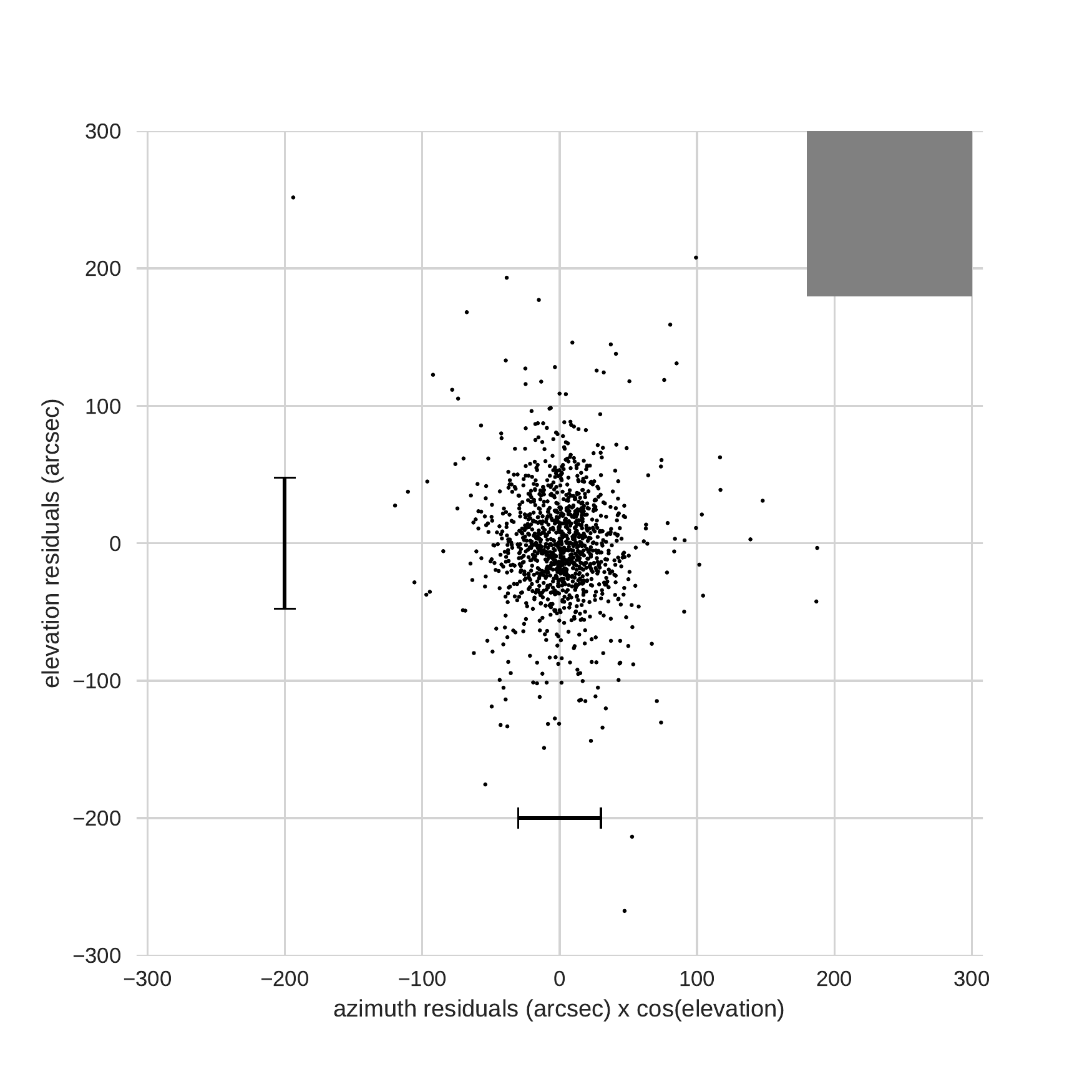}
    \caption{Residuals on the global astrometric solution for the Perenjori camera. The pixel size at the centre of the FoV is shown by the grey square in order to gauge the quality of the solution, as well as the $1\sigma$ residual bars on the stars. The azimuth residuals are artificially large around the pole of the spherical coordinate system, so we have multiplied them by $cos(elevation)$ to cancel out this artefact.}
    \label{fig:astrom}
\end{figure}

The astrometric solution for the Perenjori camera is obtained using an image taken a few hours after the event, once the clouds had cleared (2016-10-31T16:00:30\,UTC), containing 1174 stars of apparent magnitude $m_V \in [1.5, 5.5]$. A $3^{rd}$ order polynomial fit is performed to match detected stars to the Tycho-2 star catalogue. The transformation is further corrected using a $2^{nd}$ order polynomial on the radial component of the optics. The stability of the solution can be checked at regular intervals. The slight degradation in altitude precision for altitudes below \mbox{$20\degr$} in Fig. \ref{fig:astrom}, is due to a partly obstructed horizon from this camera (eg. trees, roofs). This degradation usually starts around \mbox{$10\degr$} on cameras with a clear horizon, as is the case for most outback systems.

The beginning of the fireball on the Perenjori image is partially masked by clouds, yielding only a handful of points. The middle section is not usable as the sensor was saturated in large blobs, rendering impossible timing decoding or even reliable identification of the centre of the track. However the Perenjori image provides a good viewpoint for the end of the fireball.

Well calibrated data were also obtained from the Badgingarra camera, before it went outside the sensor area at 30.6\,km altitude.
Although the Northam camera was very cloudy, we were able to pick the track of the main meteoroid body without timing information, and use it as a purely geometric constraint.
Hyden, Kukerin, and Newdegate also picked up the fireball, however the astrometry so low on the horizon ($<5\degr$ ) was too imprecise (between 2 and 4 arcminutes) to refine the trajectory solution. 

\subsection{Photometry}
The automated DFN data reduction pipeline routinely calculates brightness for non-saturated fireball segments.
For this bright event however, the brightness issue was exacerbated by large amounts of light scattered in the clouds (Fig. \ref{fig:fireball}), so it was impossible to produce a useful light curve from the photograph.
On the other hand, the Perenjori observatory recorded a low-resolution compressed video through the clouds.
Although it is not possible to calibrate this signal, we can get a remarkably deep dynamic range reading of the all-sky brightness, thanks to the large amount of light scattered in the numerous clouds. By de-interlacing the analogue video frames, we were able to effectively double the time resolution (25 interlaced frames per second to 50 fields per second, which are equally as precise for all-sky brightness measurements). 
To correct how the auto-gain affects the signal, we perform aperture photometry on Venus throughout the event.
The analogue video feed is converted to digital by the Commell MPX-885 capture card, and then processed by the compression algorithm (H264 VBR, FFmpeg \textit{ultrafast} preset) \citep{2017ExA...tmp...19H} before being written to disk, divided into 1 minute long segments. The PC clock is maintained by the Network Time Protocol (NTP) service, fed with both GNSS and network time sources. However the timestamp on the file created by the PC suffers from a delay.  We measured the average delay using a GPS video time inserter (IOTA-VTI) on a test observatory. This allowed us to match the light curve obtained from the video to astrometric data to within $20\,\mbox{ms}$.
\begin{figure}[!h]
    \centering
    \includegraphics[width=\linewidth]{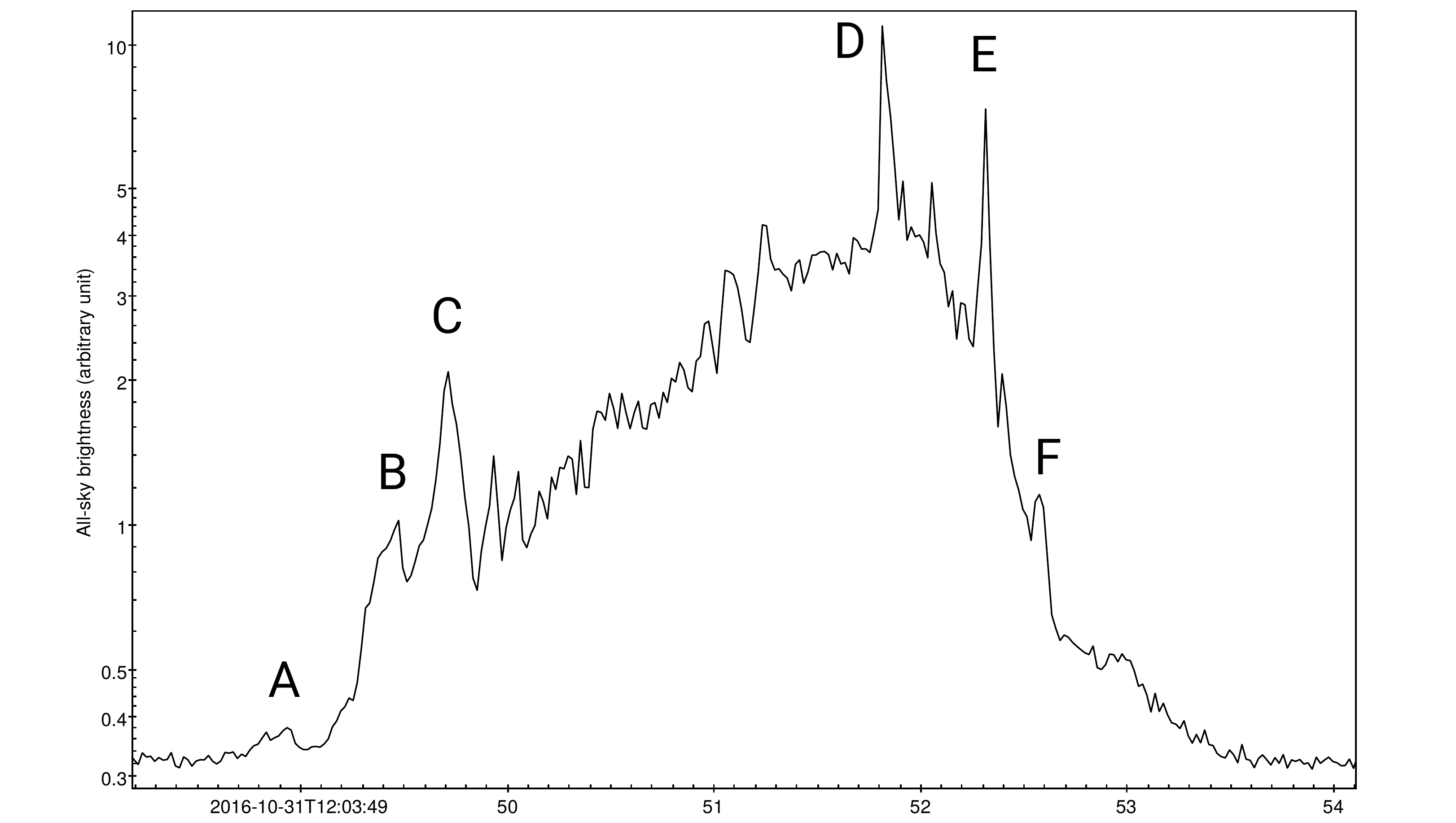}
    \caption{All-sky brightness (sum of all the pixels) from the video camera at the Perenjori observatory. The light curve is corrected to take into account the effect of auto-gain.}
    \label{fig:radiometric_light_curve}
\end{figure}
Peak \textit{A} in Figure \ref{fig:radiometric_light_curve} is visible on the photographs from both Badgingarra and Hyden. These are used to validate the absolute timing alignment of the video data.

\subsection{Eye witnesses}

Three anecdotal reports of the fireball were received via the \textit{Fireballs in the Sky} smartphone app \citep{2014LPI....45.1731P, 2016pimo.conf..267S} within two hours of the event (Table \ref{tab:fits}). The free app is designed to enable members of the public to easily report fireball sightings. Phone GPS, compass, and accelerometers are utilised to report the direction of observations, while a fireball animation aids users in estimating the colour, duration and brightness of the event. This app is an interactive alternative to the popular web-based reporting tool of the International Meteor Organisation \citep{2015JIMO...43....2H}.
\begin{table}
    \centering
    \begin{tabular}{c|c|c|c|c|c|c}
         Reporting &Report& Location & Approx. Distance & Reported    & Reported     & Reported \\
         Means     &Time  &          & From Event       & Duration (s)& Brightness   & Colour   \\
                   &(UTC)  &          & (km)             &             & (stellar Mag)&          \\
         \hline
         FITS      &12:04 & Perth    & 300              & 2.6         &  -8          & Orange   \\
                   &      & region   &                  &             &              &          \\
         FITS      &12:59 & Ballidu  & 150              & 6.4         &  -7          &  Green   \\
         FITS      &13:35 & Dowerin  & 230              & 8.6         &    -9        &  Pink    \\
         eye       &N/A   & Koolanooka & 7.4            & $> 5$       & $>-12.6 $    &        \\
         witness   &      & Hills    &                  &             & (full moon)  &          
    \end{tabular}
    \caption{Observer reports from eyewitness accounts and \textit{Fireballs in the Sky} app (FITS). }
    \label{tab:fits}
\end{table}

The app reports were the first notification of the fireball received by the DFN team, even before the receipt of daily emails from the fireball observatories. The azimuth angles reported by the observers were not sufficiently consistent to enable a triangulation based on app reports alone. 

The fireball was also reported by several nearby witnesses, and was described in detail by an eye witness only $7.4\,\mbox{km}$ from the fall position (Table \ref{tab:fits}) who also reported 
hearing sounds, which due to the time of arrival may have been electrophonic in nature \citep{keay1992electrophonic}.

\section{Fireball Trajectory Analysis}

\subsection{Geometry}\label{sec:triang}

\begin{figure}[!h]
    \centering
    \includegraphics[width=\linewidth]{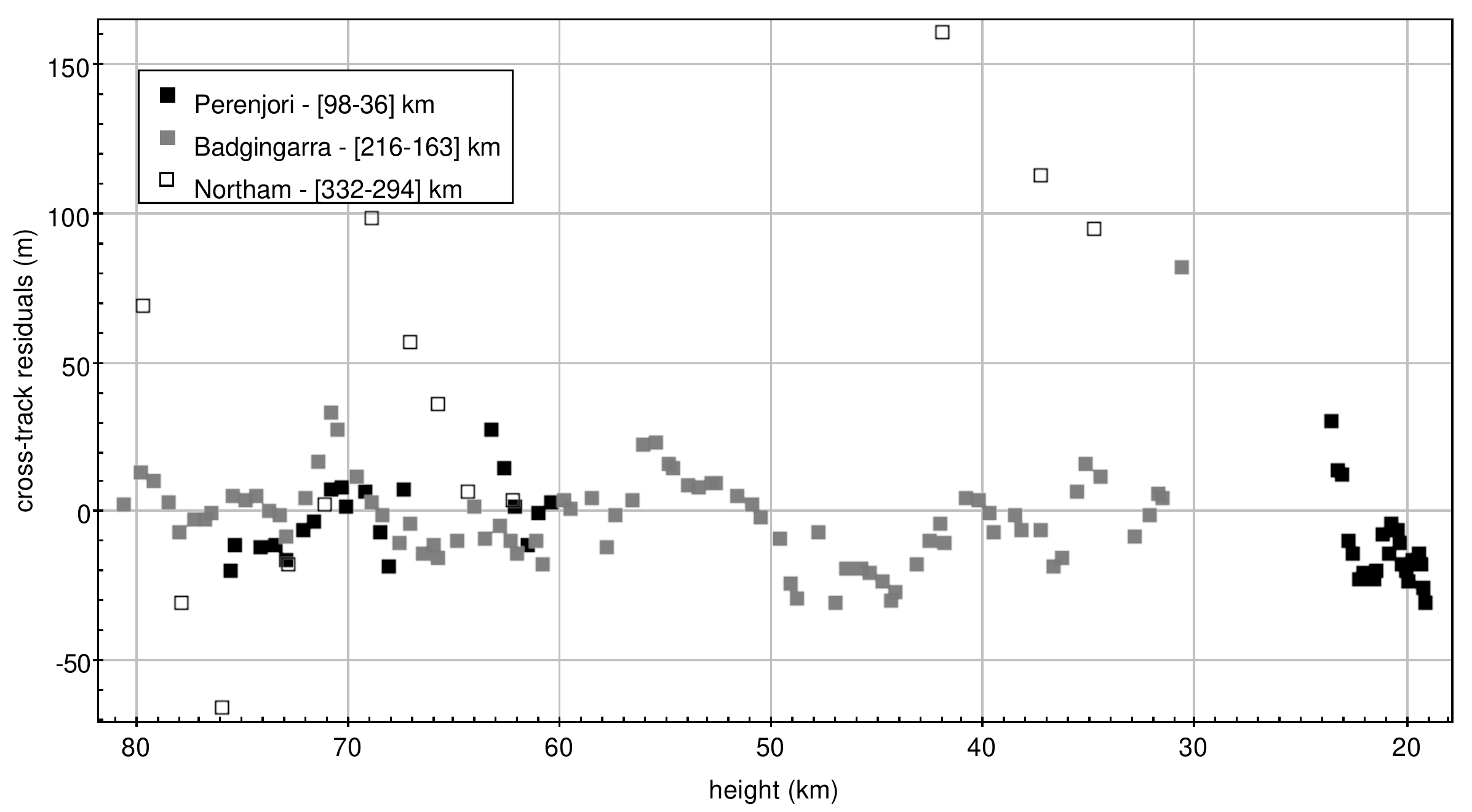}
    \caption{Cross-track residuals of the straight line least squares fit to the trajectory from each view point. These distances correspond to astrometric residuals projected on a perpendicular plane to the line of sight, positive when the line of sight falls above the trajectory solution. Note that the larger residuals on the Northam camera do not equate to larger astrometric uncertainties, but rather reflect a rather large distance from the observatory. The distances in the legend correspond to the observation range [highest point - lowest point].}
    \label{fig:cross-track_residuals}
\end{figure}

To determine the trajectory of the fireball through the atmosphere, we use a modified version of the \citet{1990BAICz..41..391B} straight-line least squares (SLLS) method. This involves creating a radiant in 3D space that best fits all the observed lines of sight, minimising the angular residuals between the radiant line and the observed lines of sight. While angular uncertainties will be similar across different camera systems, the effect of distance results in larger cross-track errors for more distant observatories (Fig. \ref{fig:cross-track_residuals}), and therefore have less influence on the resulting radiant fit.
The end of the fireball from the Perenjori image was used, along with Badgingarra and Northam camera data to triangulate the geometry of the fireball trajectory. The inclusion of astrometric data from Hyden, Kukerin, and Newdegate (see section \ref{sec:astrometry}) degraded the solution: the cross-track residuals from all viewpoints increased significantly, suggesting a systematic issue with the above mentioned camera data.
Therefore we only used the trajectory solution yielded by the 3 closest view points (Fig. \ref{fig:cross-track_residuals}).
The best combination of viewpoints (Perenjori and Badgingarra) yields an excellent convergence angle of $86\degr$.
The trajectory solution points to a moderately steep entry with a slope of $51\degr$ from the horizon, with ablation starting at an altitude of $80.6\,\mbox{km}$ and ending at $19.1\,\mbox{km}$ (see Table \ref{tab:traj_sum}).

\begin{table}[!h]
    \centering
    \begin{tabular}{lcccccc}
    \hline\\
        Event & Time\tablenotemark{*} &  Speed &  Height &  Longitude &  Latitude &  Dynamic pressure \\
         & s & $\mbox{m s}^{-1}$ & m & \degr E & \degr N & MPa \\
        \hline\\
        Beginning & 0.0 & 15443$\pm60$ & 80594 &  116.41678& -28.77573 & \\
        A & 1.20 & 15428 & 65819& 116.36429& -28.86973& 0.03   \\
        B & 1.72 & 15401 & 59444& 116.34151& -28.91045& 0.08\\
        C & 1.96 & 15378 & 56531& 116.33108& -28.92909& 0.11 \\
        D & 4.08 & 13240 & 32036& 116.24270& -29.08672& 2.28 \\
        E & 4.58 & 10508 & 27302& 116.22547& -29.11738& 3.09 \\
        F & 4.84 & 8988 & 25019& 116.21716& -29.13217& 3.27 \\
        Terminal & 6.10 & 3243 $\pm$ 465 & 19122 & 116.19564 & -29.17045 & \\
        \hline\\
    \hline
    \end{tabular}
   % \tablenotetext{+}{Fragmentation event (see \ref{fig:radiometric_light_curve)}
    \tablenotetext{*}{past 2016-10-31T12:03:47.726\,UTC}
    \caption{Summary table of bright flight events. Fragmentation event letters are defined on the light curve (Fig. \protect\ref{fig:radiometric_light_curve})}
    \label{tab:traj_sum}
\end{table}

\subsection{Dynamic modelling of the trajectory, including velocity and mass determination}\label{sec:ellie}
\paragraph{Filter Modelling}
The method described in Chapter 4 of \citet{sansom2016tracking} is an iterative Monte Carlo technique that aims to determine the path and physical characteristics such as shape ($A$: the cross section area to volume ratio), density ($\rho_m$), and ablation coefficient ($\sigma$) of a meteoroid from camera network data. 
In this approach, one is able to model meteoroid trajectories based on raw astrometric data. This avoids any preconceived constraints imposed on the trajectory, such as the straight line assumption used in Section \ref{sec:triang}. 
Unfortunately this requires multiple view points with accurate absolute timing information to record the meteoroid position.
For this event, timings encoded in the trajectory were distinguishable for only the initial $4.2$ seconds by the Badgingarra system (before any significant deceleration) and for the final 1.1 seconds by the Perenjori system. In this case we must rely on the straight-line least squares (SLLS) triangulation to determine meteoroid positions (see Section \ref{sec:triang}).
We therefore applied the three dimensional particle filter model outlined in Chapter 4 of \citet{sansom2016tracking} using instead triangulated geocentric coordinates as observation measurements. Uncertainties associated with using pre-triangulated positions based on an assumed straight line trajectory are incorporated. The distribution of particle positions using such observations will be overall greater than if we had been able to use the raw measurements. 

As a straight line may be an oversimplification of the trajectory, to most reliably triangulate the end of the luminous flight using the SLLS method, the final 1.1 seconds were isolated (this being after all major fragmentation events described in Section \ref{sec:light_curve}). The filter was run using these positions and initiated at $t_0=5.0$ seconds (2016-10-31T12:03:52.726\,UTC).
Particle mass values at this time would be more suitably initiated using a logarithmic distribution between the range of $0\,\mbox{kg}$ to $1000\,\mbox{kg}$.
The initiation of other filter parameters, including the multimodal density distribution, are described in \citet{2017AJ....153...87S} with ranges given in Table 1 of the work. 
As a calibrated light curve was not attainable, brightness values were not included in this analysis, making it a purely dynamic solution. 

\begin{figure}
    \centering
    \includegraphics[width=\linewidth]{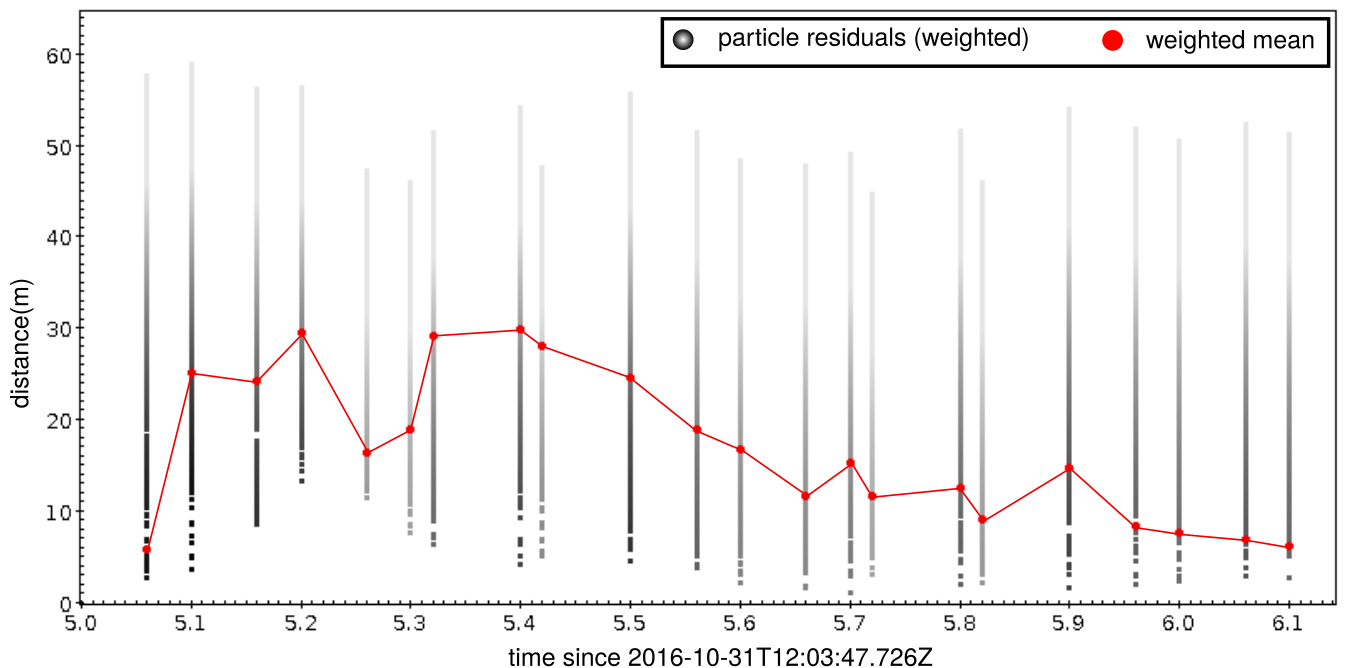}
    \caption{Position residuals of the 3D particle filter fit to the SLLS triangulated observations for the final $1.1\,s$ of the luminous trajectory. Individual particle weightings are shown in greyscales, with weighted mean values shown in red.}
    \label{fig:pf_resids}
\end{figure}

The adaptive particle filter technique applied here uses the same state vector and three dimensional state equations as in Chapter 4 of  \citet{sansom2016tracking}, to evaluate the meteoroid travelling through the atmosphere. As we are using pre-triangulated geocentric positions as observations, the measurement function here is linear. 
The particles are still allowed to move in 3D space, and an evaluation of the model fit is performed as the absolute distance between the pre-triangulated SLLS point and the evaluated particle position. This is shown in Figure \ref{fig:pf_resids} for all particles, with the distance to the mean value also shown. Mean particle positions show a good fit to the SLLS triangulated observations, with a maximum of $30\,\mbox{m}$ differences early on, decreasing to $6\,\mbox{m}$ at the end.

\begin{figure}
    \centering
    \includegraphics[width=\linewidth]{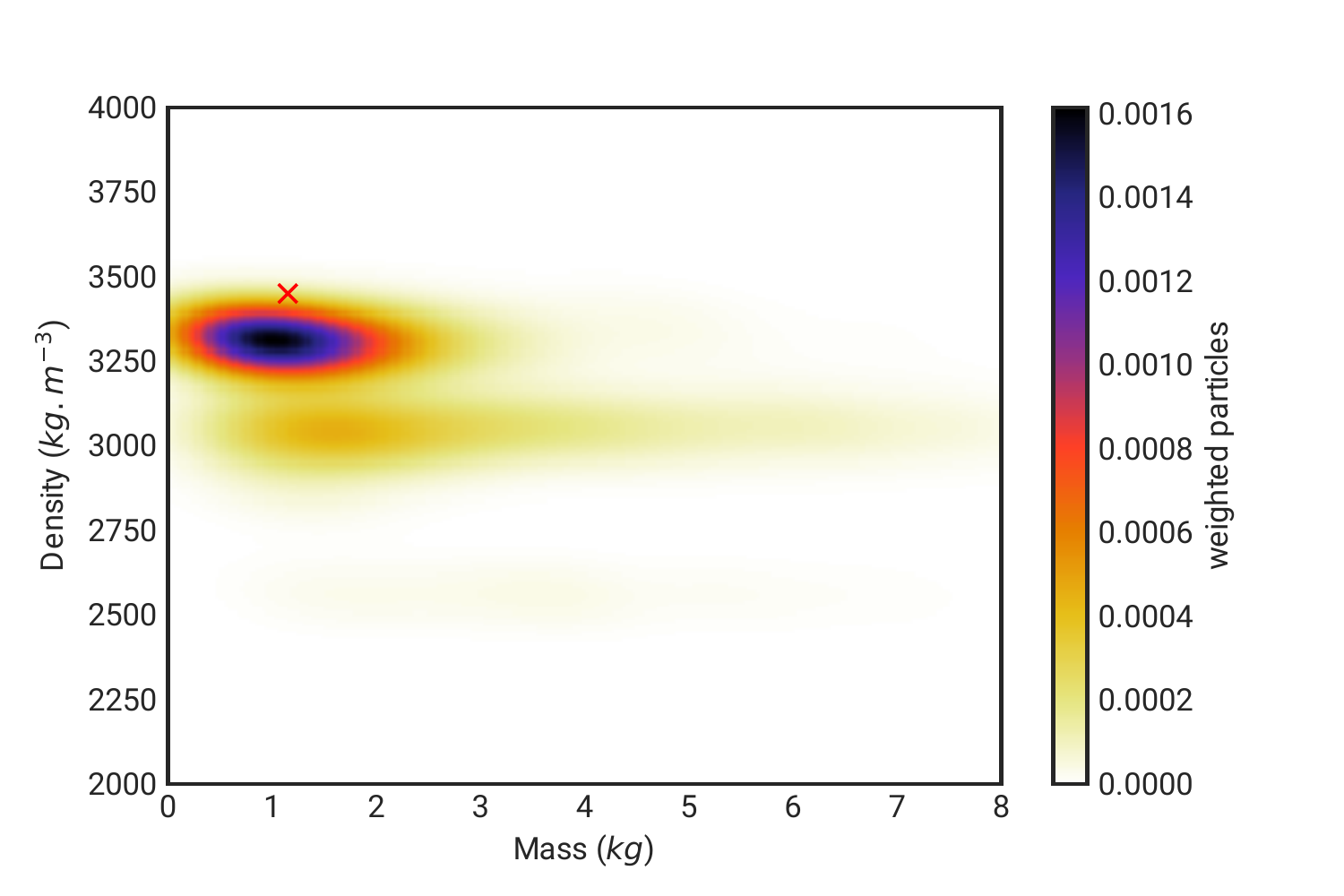}
    \caption{Results of the 3D particle filter modelling, showing the distribution of final mass estimates along with the densities with which particles were initiated at $t_0=5\,\mbox{s}$. Mass estimates are consistent with the recovered meteorite mass found (red cross), with initial densities slightly below the bulk rock value. }
    \label{fig:mass_distribution_pf_analysis}
\end{figure}

\setcounter{footnote}{0} % stupid footnotes... won't let me just have a damn *
The filter estimates not only the position and velocity of the meteoroid at each observation time, but also the mass, ablation coefficient, $\sigma$, and shape density coefficient, $\kappa$.
At the final observation time $t_f =6.1\,\mbox{s}$ (2016-10-31T12:03:53.826\,UTC), the state estimate results in weighted median values of 
$mass_f     = 1.49   \pm 0.23\,\mbox{kg}$, 
$speed_f = 3359   \pm  72   \,\mbox{m s}^{-1}$, 
$\sigma_f   = 0.0154 \pm  0.0054   \, s^2\,\mbox{km}^{-2}  $ and 
$\kappa_f  = 0.0027 \pm  0.0001  \,\mbox{(SI)}$. 
Although $\kappa$ may be used to calculate densities for a given shape and drag coefficient, to avoid introducing assumptions at this stage we may gauge its value by reviewing the density with which surviving particles were initiated. 
The distribution of final mass estimates is plotted against this initial density attributed to each given particle in Figure \ref{fig:mass_distribution_pf_analysis}, along with the recovered Dingle Dell meteorite mass of $1.150\,\mbox{kg}$ and bulk density of $3450\,\mbox{kg m}^{-3}$.
In this figure, the distribution of the main cluster of particles is consistent with the recovered mass, however the initial densities are lower. The weighted median value of initial bulk densities (at $t_0=5.0$\,s) for all particles re-sampled at $t_f$ is $ 3306\,\mbox{kg m}^{-3} $. 
It is expected that the bulk density of a meteoroid body may slightly increase throughout the trajectory as lower density, more friable material is preferentially lost. This could justify the slightly lower bulk densities attributed at $t_0$. 

In order to obtain the entry speed of the meteoroid with appropriate errors, we apply an extended Kalman smoother \citep{2015M&PS...50.1423S} to the straight line solution for the geometry, considering the timing of the points independently for each observatory.
Of the two cameras that have timing data for the beginning of the trajectory, only Badgingarra caught the start, giving an entry speed of $15402\pm60$\,$\mbox{m s}^{-1}$ ($1\sigma$) at 80596\,m altitude. To determine whether speeds calculated are consistent between observatories, the first speed calculated for Perenjori -- $15384\pm64$\,$\mbox{m s}^{-1}$ at 75548\,m altiude -- is compared to the Badgingarra solution at this same altitude --$15386\pm43\,\mbox{m s}^{-1}$. The results are remarkably consistent, validating the use of a Kalman smoother for determining initial velocities.

\begin{figure}
    \centering
    \includegraphics[width=\linewidth]{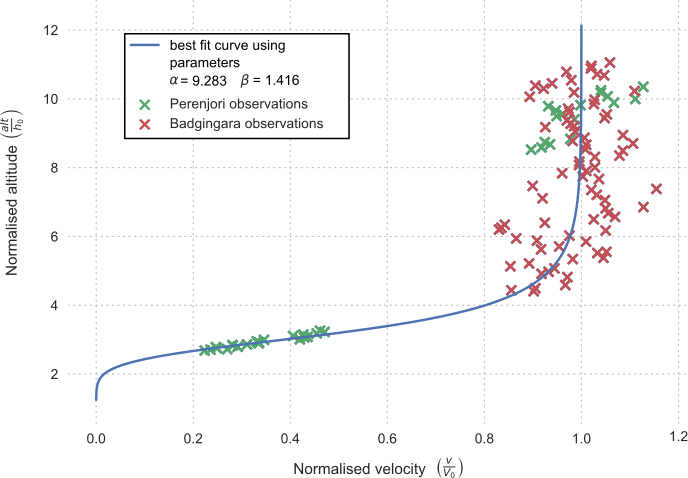}
    \caption{Trajectory data from both Perenjori and Badgingarra observatories, with speeds normalised to the speed at the top of the atmosphere (15.443\,$\mbox{km s}^{-1}$; Tab. \ref{tab:traj_sum}), $V_0$, and altitudes normalised to the atmospheric scale height, $h_0 = 7.16\,\mbox{km}$. The best fit to Equation 10 of \citet{2009AdSpR..44..323G} results in  $\alpha=9.283$ and $\beta=1.416$ and is shown by the blue line. These dimensionless parameters can be used to determine the entry and terminal mass of the Dingle Dell meteoroid.}
    \label{fig:ball}
\end{figure}

\paragraph{Dimensionless Coefficient Method}\label{sec:gritsevich}
As a comparison to the particle filter method, the dimensionless parameter technique described by \citet{2009AdSpR..44..323G} was also applied. 
The ballistic parameter ($\alpha$) and the mass loss parameter ($\beta$) were calculated for the event, resulting in $\alpha = 9.283$ and $\beta = 1.416$ (Figure \ref{fig:ball}). 
As the particle filter technique in this case was not able to be performed on the first 5.0 seconds of the luminous trajectory, these parameters may be used to determine both initial\footnote{see equation 14 in \citet{2009AdSpR..44..323G}}, and final\footnote{see equation 6 in \citet{2009AdSpR..44..323G}} main masses, given assumed values of the shape and density of the body. Using the same parameters as \citet{2009AdSpR..44..323G} ($c_d = 1$, $A = 1.55$) along with the density of the recovered meteorite, $\rho = 3450\,\mbox{kg m}^{-3}$, gives an entry mass, $m_e = 81.6\,\mbox{kg}$, and a $m_f=1.4\,\mbox{kg}$. Varying the shape of the body to spherical values, $A=1.21$ \citep{Bronshten1983} gives an initial mass of $m_e=38.8\,\mbox{kg}$. Instead of assuming values for $c_d$ and $A$, we can also insert the $\kappa$ value calculated by the particle filter to give $m_e=41.1\,\mbox{kg}$.
These results can be approximated to a 30\,cm diameter initial body.
Note that this method is the most reliable for calculating a minimum entry mass of the Dingle Dell meteoroid. The photometric method would require a calibrated light curve, and the particle filter method requires good astrometric data coverage where significant deceleration occurs (the missing data between $4.2$ and $5.0$ seconds).

\subsection{Atmospheric behaviour}\label{sec:light_curve}

In Table \ref{tab:traj_sum} we report the ram pressure ($P = \rho_a v^2$) required to initiate the major fragmentation events labelled on the light curve in Fig. \ref{fig:radiometric_light_curve}. The density of the atmosphere,  $\rho_a$, is calculated using the \textit{NRLMSISE-00} model of \citet{2002JGRA..107.1468P}, and $v$ is the calculated speed.
The meteoroid started fragmenting quite early (events \textit{A}, \textit{B}, and \textit{C}), starting at $0.03\,$MPa. These early fragmentation events suggest that the meteoroid had a much weaker lithology than the meteorite that was recovered on the ground.
Then no major fragmentation happened until two very bright peaks in the light curve: \textit{D} ($2.28\,$MPa) and \textit{E} ($3.09\,$MPa). These large short-lived peaks suggest a release of a large number of small pieces that quickly burnt up. A small final flare (\textit{F}--$3.27\,$MPa) 1.26 second before the end is also noted.

\section{Dark Flight and Meteorite Recovery}

The results of the dynamic modelling (Fig. \ref{fig:mass_distribution_pf_analysis}) are fed directly into the dark flight routine. By using the state vectors (both dynamical and physical parameters) from the cloud of possible particles, we ensure that there is no discontinuity between the bright flight and the dark flight, and we get a simulation of possible impact points on the ground that is representative of the modelling work done on bright flight data.

\subsection{Wind modelling}\label{sec:wind}
\begin{figure}[!h]
    \centering
    \includegraphics[width=\linewidth]{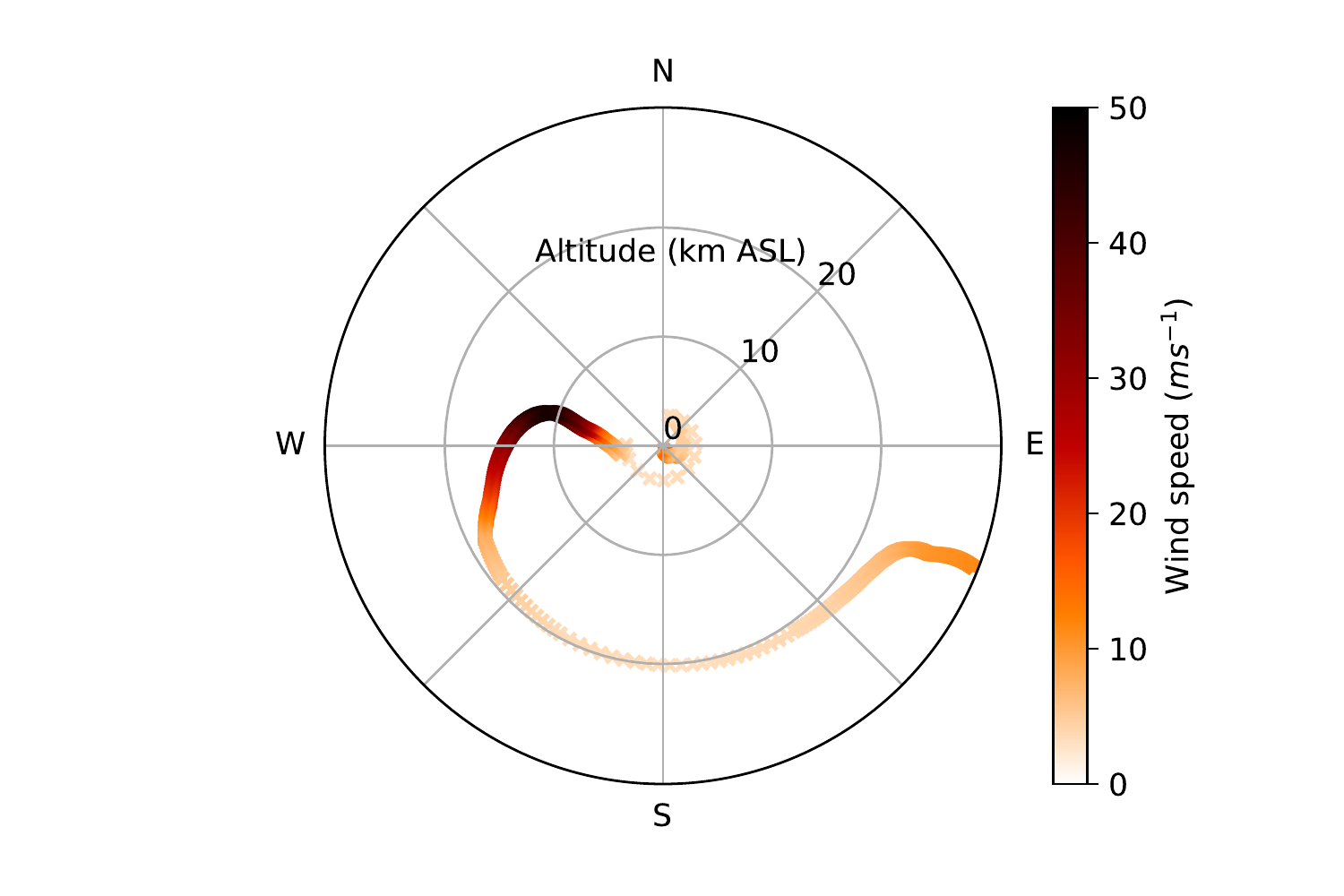}
    \caption{Wind model profile \textit{W1}, extracted as a vertical profile at the coordinates of the lowest visible bright flight measurement.}
    \label{fig:wind_model}
\end{figure}

The atmospheric winds were numerically modelled using the Weather Research and Forecasting (WRF) software package version 3.8.1 with the Advanced Research WRF (ARW) dynamic solver \citep{skamarock2008description}. The weather modelling was initialised using global $1\degr$ resolution National Centers for Environmental Prediction (NCEP) Final analysis (FNL) Operational Model Global Tropospheric Analysis data. As a result, a 3\,km resolution WRF product with 30 minutes history interval was created and weather profile at the end of the luminous flight for 2016-10-31T12:00\,UTC extracted (Fig. \ref{fig:wind_model}). The weather profile includes wind speed, wind direction, pressure, temperature and relative humidity at heights ranging up to 30\,km (Fig. \ref{fig:wind_model}), providing complete atmospheric data for the main mass from the end of the luminous phase to the ground, as well as for fragmentation events \textit{E} and \textit{F} (Table \ref{tab:traj_sum}).
Different wind profiles have been generated, by starting the WRF integration at different times: 2016 October 30d12h, 30d18h, 31d00h, 31d06h, and 31d12h UTC.
Three of the resulting wind models converge to a similar solution in both speed and direction (30d12h, 31d00h, 31d12h) and will be hereafter referred to as solution \textit{W1} (Fig. \ref{fig:wind_model}).
The other two models from 30d18h (\textit{W2}) and 31d00h (\textit{W3}) differ significantly. For example, the maximum jet stream strength is $\simeq47\,\mbox{m s}^{-1}$ for \textit{W1}, $\simeq34\,\mbox{m s}^{-1}$ for \textit{W3}, and $\simeq29\,\mbox{m s}^{-1}$ for \textit{W2}.
To discriminate which wind profile is closer to the truth, we ran the model next to the Geraldton balloon launches of 2016 October 31d00h and 31d06h UTC, but no discrepancy was noticeable between all 5 scenarios.
Considering that 3 model runs clump around \textit{W1}, whereas \textit{W3} and \textit{W2} are isolated, we choose \textit{W1} as a preferred solution.  The investigation of why \textit{W3} and \textit{W2} are different is beyond the scope of this paper, nonetheless we discuss how these differences affect the dark flight of the meteorites in the next section (\ref{sec:df}).

\subsection{Dark flight integration}\label{sec:df}
The calculations of the unobserved terminal part of the ablation phase and the dark flight are performed using an $8^{th}$ order explicit Runge-Kutta integrator with adaptive step-size control for error handling. The physical model uses the single body equations for meteoroid deceleration and ablation \citep{hoppe1937physikalischen, 1939PAAS....9R.136W}. 
In this model, rotation is accounted for such that the cross sectional area to volume ratio ($A$) remains constant throughout the trajectory.
The variation in flow regimes and Mach ranges passed through by the body alter the values used for the drag coefficient, which can be approximated using Table\,1 in \citep{2015M&PS...50.1423S}.

The integration of all the particles from Section \ref{sec:ellie} allows the generation of probability heat maps to maximise field searching efficiency.
The ground impact speed for the mass corresponding to the recovered meteorite is evaluated at $67\,\mbox{m s}^{-1}$.

In calculating a fall line for an arbitrary range of masses, the assumed shape of the body and the wind model used both affect the final fall position. However for a given wind model a change in shape only shifts the masses along the fall line.

\begin{figure}[!h]
    \centering
    \includegraphics[width=\linewidth]{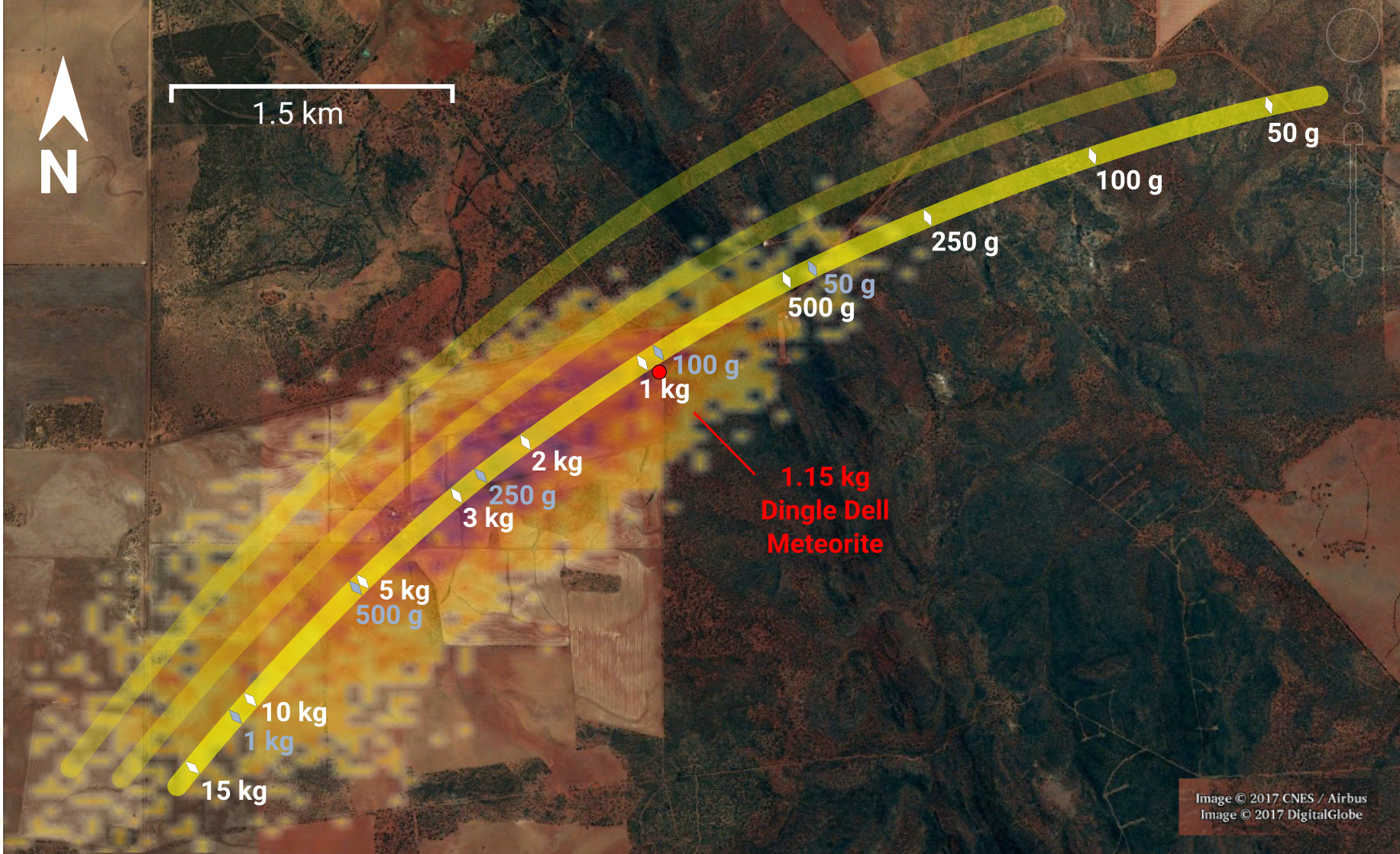}
    \caption{Fall area around Dingle Dell farm and Koolanooka Hills. Fall lines in yellow represent different wind model solutions: \textit{W1} (bottom), \textit{W2} (middle) and \textit{W3} (top). Mass predictions for the preferred wind model are shown for spherical (light blue markings; $A=1.21$) and cylindrical (white markings; $A=1.5$) assumptions. The particle filter results are propagated through dark flight using wind model \textit{W1}, %assuming a brick shape (values given in Section \ref{sec:ellie}), 
    and are shown as a heat map. The location of the recovered meteorite (red dot) is $\simeq 100\,\mbox{m}$ from the \textit{W1} fall line.}
    \label{fig:fall_line}
\end{figure}

We also calculate dark flight fall lines from fragmentation events that happened within the wind model domain: \textit{E} and \textit{F}.
Unsurprisingly, the main masses from those events are a close match to the corresponding main mass started from the end of the visible bright flight.
However small fragments are unlikely to be found as they fell into the Koolanooka Hills bush land (Fig. \ref{fig:fall_line}).

\subsection{Search and recovery}

\begin{figure}[!h]
    \centering
    \includegraphics[width=\linewidth]{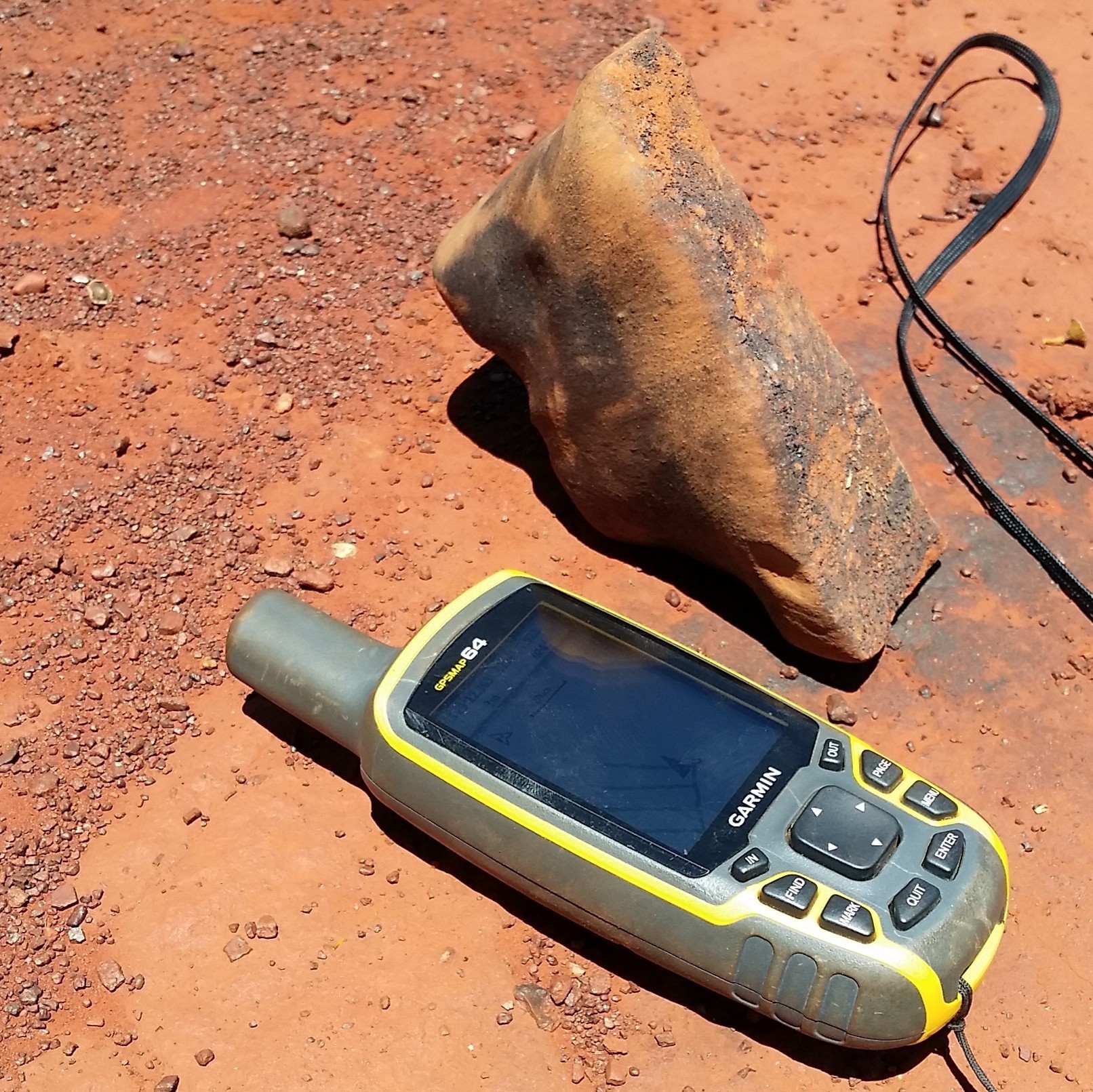}
    \caption{Dingle Dell meteorite as it was found. Image available at \url{https://commons.wikimedia.org/wiki/File:Dingle_Dell_meteorite_as_it_was_found.jpg} under a Creative Commons Attribution-ShareAlike 4.0 International.}
    %Searching pattern is visible on the GPS unit.
    \label{fig:rock}
\end{figure}

Within two days, two of the authors (PB and MT) visited the predicted fall area, about 4 hours' drive from Perth, Western Australia to canvas local farmers for access and information. 
Having gained landowner permission to search, a team was sent to the area 3 days later.
Searching was carried out by a team of 4 (MT, BH, TJS, and HD), mostly on foot and with some use of mountain biking in open fields. The open fields' searching conditions were excellent, although the field boundaries were vegetated. The team managed to cover about 12\,ha per hour when looking for a $>1\,\mbox{kg}$ mass on foot. On the second day, a meteorite was found (Fig. \ref{fig:rock}) close to the Dingle Dell farm boundary,
 at coordinates $\lambda = 116.215439\degr$ $\phi = -29.206106\degr$ (WGS84), about $130\,\mbox{m}$ from the originally calculated fall line, after a total of 8 hours of searching.
 The recovered meteorite weighs 1.15\,kg, with a rounded brick shape of approximately 16\,x\,9\,x\,4\,cm, and a calculated bulk density of $3450\,\mbox{kg m}^{-3}$ (Fig. \ref{fig:rock}). The condition of the meteorite is excellent, having only been on ground for 6 days, 16 hours.
 Discussion with the local landowner, and checking the weather on the nearest Bureau Of Meteorology observation station (Morawa Airport, $20\,\mbox{km}$ away) showed that no precipitation had fallen between times of landing and recovery. The meteorite was collected and stored using a Teflon bag, and local soil samples were also collected in the same manner for comparison.
No trace of impact on the ground was noticed. The meteorite was found intact (entirely covered by fusion crust) on hard ground, resting up-right (Fig. \ref{fig:rock}), and covered with dust.
So it is possible that the meteorite fell a few metres away in softer ground and bounced or rolled to the recovered position.

 %Some fine red dust was seen on surface, most probably aeolian in nature. The ground where the meteorite was found consisted of hard laterite; but examination of the meteorite shows no sign of crushing/fracture on any surfaces from a landing impact, so it is possible that the meteorite fell a few metres away in softer ground and bounced or rolled to the recovered position. This might also explain the fine dust coating on the surface. \textbf{ goes too far in the rock decription??}

\section{Pre-encounter orbit}

The backward propagation of the observed trajectory into an orbit requires the calculation of the direction of the fireball (known as the radiant), and the position and speed at the top of the atmosphere. The associated uncertainties on these two components are mostly un-correlated.
In order to minimise issues associated with the oversimplified straight line trajectory for orbit purposes, we re-triangulate the observations using only points that fall $>60\,\mbox{km}$ altitude on the initial triangulation.
In this case, as the trajectory is fairly steep, the difference in apparent radiant between the two solutions is less than 5\,arcmin.
To calculate the errors on the radiant, we use the co-variance matrix from the least squares trajectory fit (see section \ref{sec:triang}), this gives us the apparent radiant: 
slope to the horizontal = $51.562\pm 0.002\degr$, azimuth of the radiant (East of North) = $26.17 \pm 0.03\degr$, which corresponds to ($\alpha = 353.38 \pm 0.02\degr$, $\delta = 6.34 \pm 0.01\degr$)  in equatorial J2000 coordinates.

To calculate the formal uncertainty on the initial velocity, we apply the Kalman filter methods of \citet{2015M&PS...50.1423S} as outlined in Section \ref{sec:ellie}.
Using the time, position, radiant, speed, and their associated uncertainties, we determine the pre-atmospheric orbit by propagating the meteoroid trajectory back through time, considering the atmospheric deceleration, Earth's oblate shape effects (J2), and other major perturbing bodies (such as the Moon and planets), until the meteoroid has gone beyond 10$\times$ the Earth's sphere of influence. From here, the meteoroid is propagated forward in time to the impact epoch, ignoring the effects of the Earth-Moon system. Uncertainties (Table \ref{tab:orbit}) are calculated using a Monte Carlo approach on 1000 test particles randomly drawn using uncertainties on the radiant and the speed.

\begin{figure}[!h]
    \centering
    \includegraphics[width=\linewidth]{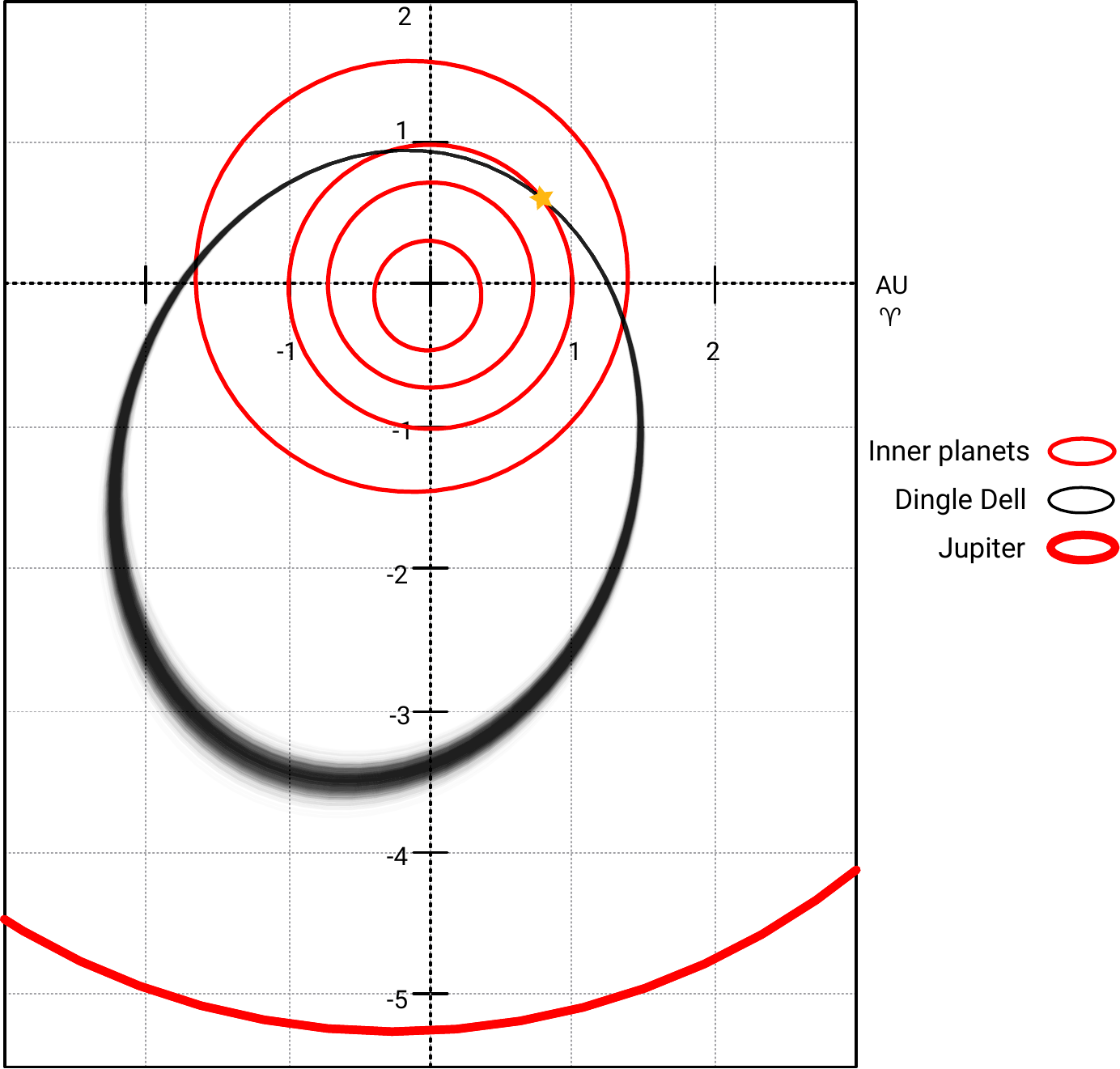}
    \caption{Ecliptic projection of the pre-encounter orbit of Dingle Dell. The shades of grey represent the likelihood as calculated from 1000 Monte Carlo simulations based on formal uncertainties on the radiant and the speed.}
    \label{fig:dd_orbit}
\end{figure}

\begin{table}[!h]
    \centering
    \begin{tabular}{ccc}
    \hline 
    Epoch & TDB & 2016-10-31\\
         $a$ & AU & 2.254 $\pm$ 0.034\\
  $e$ & &  0.5904 $\pm$ 0.0063\\
  $i$ & \degr   & 4.051 $\pm$ 0.012\\
   $\omega$ & \degr  & 215.773$\pm$ 0.049\\
    $\Omega$ & \degr  &  218.252 $\pm$ 0.00032\\
     $q$ & AU &  0.92328 $\pm$ 0.00032\\
     $Q$ & AU & 3.586 $\pm$ 0.067\\
     $\alpha _g$ & \degr   & 354.581 $\pm$ 0.037\\
     $\delta _g$ & \degr  & 13.093 $\pm$ 0.081\\
     $V _g$ & $\mbox{m s}^{-1}$ & 10508 $\pm$ 87\\
     $T _J$ & & 3.37 \\ %5.20336301/2.25+2*cos(4.051 degrees)*sqrt(2.25/5.20336301*(1-0.5904^2))
    \hline 
    \end{tabular}
    \caption{Pre-encounter orbital parameters expressed in the heliocentric ecliptic frame (\textit{J2000}) and associated $1\sigma$ formal uncertainties.}
    \label{tab:orbit}
\end{table}

We scanned the \textit{Astorb}\footnote{\url{ftp://ftp.lowell.edu/pub/elgb/astorb.html}, downloaded June 24, 2017} asteroid orbital database \citep{2002aste.book...27B} for close matches in $a,e,i,\omega,\Omega$ orbital space using the similarity criterion of \citet{1963SCoA....7..261S}.
The closest match is the small ($H=24.6$) \textit{2015\,TD179} asteroid, that came into light in November 2015 when it flew by Earth at $\simeq10$\,lunar distances. But the large difference between these orbits, $D=0.04$, makes the dynamical connection between the two highly unlikely.

To calculate the likely source region and dynamical pathway that put the meteoroid on an Earth crossing orbit, we use the \textit{Rebound} integrator \citep{2015MNRAS.452..376R} to backward propagate the orbit of the meteoroid.
We use 10,000 test particles randomly selected using the radiant and speed uncertainties as explained above, as well as the major perturbating bodies (Sun, 8 planets, and Moon).
The initial semi-major axis (Table \ref{tab:orbit}) is close to the 7:2 (2.25\,AU) and 10:3 (2.33\,AU) mean motion resonances with Jupiter (MMRJ). These minor resonances start to scatter the eccentricity of a large number of test particles very early on, but neither are strong enough to decrease it significantly enough to take the meteoroid outside of Mars' orbit.
Because of the interactions with the inner planets, the particle cloud rapidly spreads out, and particles gradually start falling into the two main dynamical pathways in this region: 3:1 MMRJ (2.5\,AU) and the $\nu_6$ secular resonance. These resonances rapidly expand the perihelia of particles out of the Earth's orbit initially, and eventually out of Mars' orbit and into the main belt.

During the integration over 1 million years, we count the number of particles that have converged close to stably populated regions of the main belt, and note which dynamical pathway they used to get there. This gives us the following statistics:
\begin{itemize}
\item $\nu_6$: 12\%
\item 3:1 MMRJ: 82\%
\item 5:2 MMRJ: 6\%
\end{itemize}

\begin{figure}[!h]
    \centering
    \includegraphics[width=\linewidth]{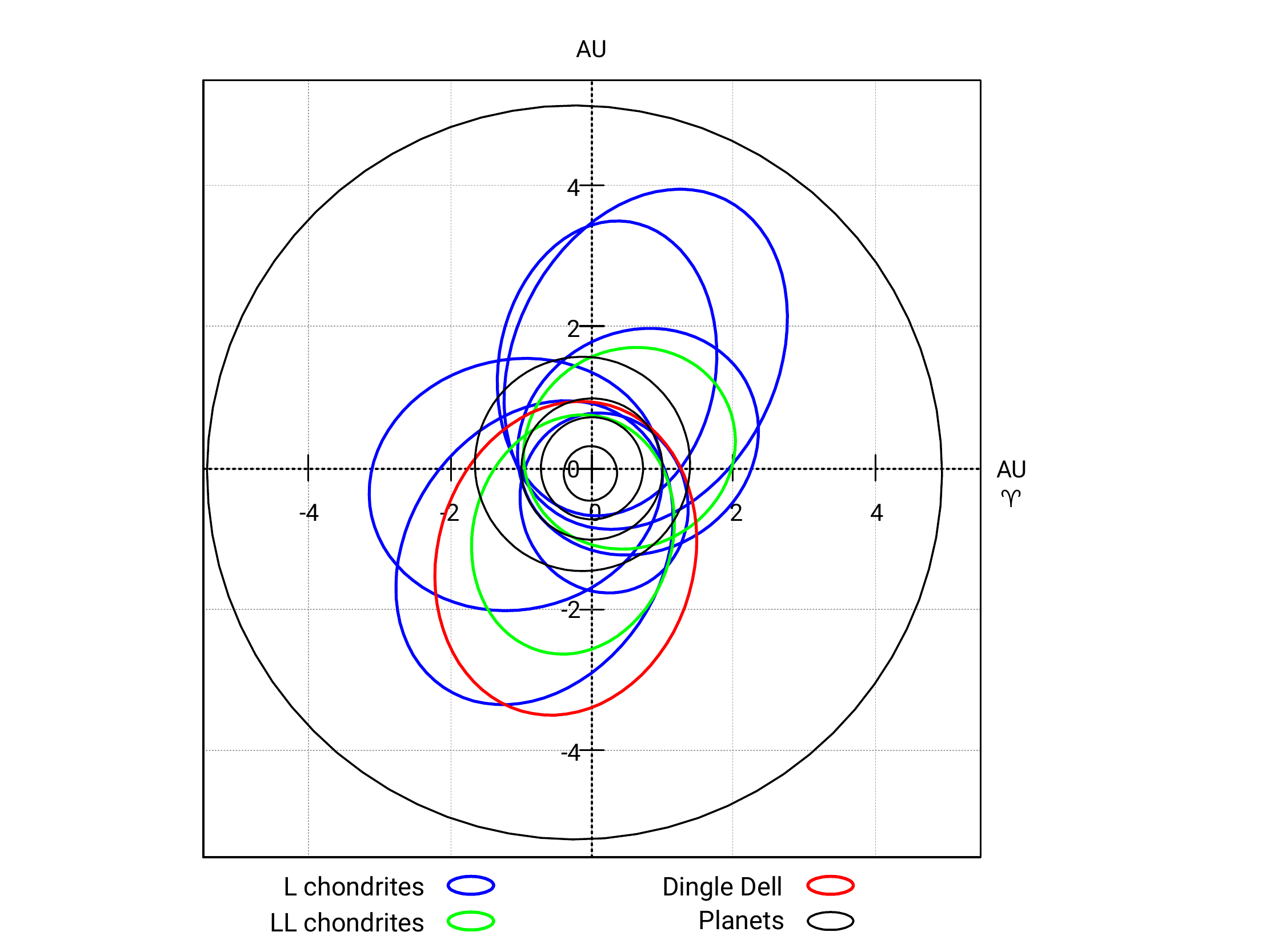}
    \caption{The orbit of Dingle Dell in context with other L and LL ordinary chondrite falls. References for L and LL orbits are in the introduction.}
    \label{fig:orb_DD_L_LL}
\end{figure}

\section{Conclusions}

Dingle Dell is the fourth meteorite with an orbit recovered by the DFN in Australia.
Its luminous trajectory was observed by 6 DFN camera stations up to 535\,km away at 12:03:47.726 UTC on 31 October, 2016. 
Clouds severely affected the observations, but enough data was available to constrain the search area for a swift recovery, and determine one of the most precise orbits linked to a meteorite.
The surviving rock was recovered within a week of its fall, without any precipitation contaminating the rock, confirming the DFN as a proficient sample recovery tool for planetary science.
This recovery, in less than ideal conditions, also validates various choices in the design and operations of the Desert Fireball Network:
\begin{itemize}
    \item Use of high resolution digital cameras to enable reliable all-sky astrometry for events up to $300\,\mbox{km}$ away.
    \item Uninterrupted operation even when a large portion of the sky is cloudy for individual systems.
    \item Archiving of all raw data to mitigate event detection failures.
\end{itemize}

While the method of \citet{2017AJ....153...87S} was still in development at the time of the fall, the re-analysis of the fireball with this new technique is remarkably consistent with the main mass found, requiring just a small number of high quality astrometric data points. This validates the method, and will drastically reduce the search area for future observed falls.

After a 1 million year integration of 10,000 test particles, it is most likely that Dingle Dell was ejected from the main belt through the 3:1 mean motion resonance with Jupiter rather than the $\nu_6$ resonance (82\% for the 3:1 MMRJ compared to 12\% probability for $\nu_6$).
This also means that L/LL Dingle Dell is unlikely to be associated with the Flora family, likely source of most LL chondrites \citep{2008Natur.454..858V}, as the most efficient mechanism for getting Florian fragments to near-Earth space is the $\nu_6$ secular resonance.
This fall adds little insight into the Flora/LL link, but 2016 was rich in instrumentally observed LL falls, which might yield clues to help confirm this connection in the near future: Stubenberg (LL6) \citep{2016LPICo1921.6221S, 2017M&PS...52.1683B}, Hradec Kr\'alov\'e (LL5) \citep{Metbull_106}, and Dishchii'bikoh (LL7) \citep{Metbull_106,2018arXiv180105072P}.

\section*{acknowledgements}
The authors would like to thank the people hosting the observatories, members of the public reporting their sightings through the \textit{Fireballs In The Sky} program, and other volunteers that have made this project possible. This research was supported by the Australian Research Council through the Australian Laureate Fellowships scheme and receives institutional support from Curtin University.
The DFN data reduction pipeline makes intensive use of Astropy, a community-developed core Python package for Astronomy \citep{2013A&A...558A..33A}.
The authors would also like to thank J. Borovi{\v c}ka and J. Vaubaillon for their valuable comments and suggestions, which improved the quality of the paper.

\section*{Appendix A. Supplementary files}
We provide the raw astrometric tables for the 3 cameras used for computing the trajectory.

We also give the straight line trajectory solution (latitude, longitude, height), as well as the corresponding speeds calculated by the method of \citet{2015M&PS...50.1423S} using all the data available (this explains slight differences with the manuscript, as in the latter they were calculated separately for each camera).

Note that the number of decimals given in these tables is not necessarily representative of uncertainty.

To illustrate the meteorite searching strategy we provide the GPS tracks, as every member of the search team carried a GPS unit (see Fig. \ref{fig:rock}). Note that one GPS unit malfunctioned, this resulted in the loss of one of the tracks on the first afternoon of the search, and explains apparent gaps in the searching grid.

We include the preferred weather model used for dark flight integration (\textit{W1}, from Fig. \ref{fig:wind_model}), extracted as a vertical profile.

\bibliography{biblio}

\begin{thebibliography}{61}
\providecommand{\natexlab}[1]{#1}
\providecommand{\url}[1]{\texttt{#1}}
\expandafter\ifx\csname urlstyle\endcsname\relax
  \providecommand{\doi}[1]{doi: #1}\else
  \providecommand{\doi}{doi: \begingroup \urlstyle{rm}\Url}\fi

\bibitem[Met(2015)]{Metbull_104}
\emph{Meteoritical Bulletin}, 104, 2015.

\bibitem[Met(2016)]{Metbull_105}
\emph{Meteoritical Bulletin}, 105, 2016.

\bibitem[Met(2017)]{Metbull_106}
\emph{Meteoritical Bulletin}, 106, 2017.

\bibitem[{Astropy Collaboration} et~al.(2013){Astropy Collaboration},
  {Robitaille}, {Tollerud}, {Greenfield}, {Droettboom}, {Bray}, {Aldcroft},
  {Davis}, {Ginsburg}, {Price-Whelan}, {Kerzendorf}, {Conley}, {Crighton},
  {Barbary}, {Muna}, {Ferguson}, {Grollier}, {Parikh}, {Nair}, {Unther},
  {Deil}, {Woillez}, {Conseil}, {Kramer}, {Turner}, {Singer}, {Fox}, {Weaver},
  {Zabalza}, {Edwards}, {Azalee Bostroem}, {Burke}, {Casey}, {Crawford},
  {Dencheva}, {Ely}, {Jenness}, {Labrie}, {Lim}, {Pierfederici}, {Pontzen},
  {Ptak}, {Refsdal}, {Servillat}, and {Streicher}]{2013A&A...558A..33A}
{Astropy Collaboration}, T.~P. {Robitaille}, E.~J. {Tollerud}, P.~{Greenfield},
  M.~{Droettboom}, E.~{Bray}, T.~{Aldcroft}, M.~{Davis}, A.~{Ginsburg}, A.~M.
  {Price-Whelan}, W.~E. {Kerzendorf}, A.~{Conley}, N.~{Crighton}, K.~{Barbary},
  D.~{Muna}, H.~{Ferguson}, F.~{Grollier}, M.~M. {Parikh}, P.~H. {Nair}, H.~M.
  {Unther}, C.~{Deil}, J.~{Woillez}, S.~{Conseil}, R.~{Kramer}, J.~E.~H.
  {Turner}, L.~{Singer}, R.~{Fox}, B.~A. {Weaver}, V.~{Zabalza}, Z.~I.
  {Edwards}, K.~{Azalee Bostroem}, D.~J. {Burke}, A.~R. {Casey}, S.~M.
  {Crawford}, N.~{Dencheva}, J.~{Ely}, T.~{Jenness}, K.~{Labrie}, P.~L. {Lim},
  F.~{Pierfederici}, A.~{Pontzen}, A.~{Ptak}, B.~{Refsdal}, M.~{Servillat}, and
  O.~{Streicher}.
\newblock {Astropy: A community Python package for astronomy}.
\newblock \emph{\aap}, 558:\penalty0 A33, Oct. 2013.
\newblock \doi{10.1051/0004-6361/201322068}.

\bibitem[{Benedix} et~al.(2017){Benedix}, {Forman}, {Daly}, {Godel}, {Esteban},
  {Meier}, {Maden}, {Busemann}, {Yin}, {Sanborn}, {Ziegler}, {Strasser},
  {Glavin}, {Dworkin}, {Bland}, {Paxman}, {Towner}, {Cup\'ak}, {Sansom},
  {Howie}, {Devillepoix}, {Cox}, {Jansen-Sturgeon}, {Hartig}, and
  {Bevan}]{benedix2017dd}
G.~K. {Benedix}, L.~V. {Forman}, L.~{Daly}, B.~{Godel}, L.~{Esteban}, M.~M.~M.
  {Meier}, C.~{Maden}, H.~{Busemann}, Q.~Z. {Yin}, M.~{Sanborn}, K.~{Ziegler},
  W.~K.~C. C. M.~W. {Strasser}, J.~W., D.~P. {Glavin}, J.~P. {Dworkin}, P.~A.
  {Bland}, J.~P. {Paxman}, M.~C. {Towner}, M.~{Cup\'ak}, E.~K. {Sansom}, R.~M.
  {Howie}, H.~A.~R. {Devillepoix}, M.~A. {Cox}, T.~{Jansen-Sturgeon}, B.~A.~D.
  {Hartig}, and A.~W.~R. {Bevan}.
\newblock {Mineralogy, Petrology and Chronology of the Dingle Dell Meteorite}.
\newblock In \emph{80th Annual Meeting of the Meteoritical Society 2017}, page
  6229, 2017.

\bibitem[{Bischoff} et~al.(2017){Bischoff}, {Barrat}, {Bauer}, {Burkhardt},
  {Busemann}, {Ebert}, {Gonsior}, {Hakenm{\"u}ller}, {Haloda}, {Harries},
  {Heinlein}, {Hiesinger}, {Hochleitner}, {Hoffmann}, {Kaliwoda},
  {Laubenstein}, {Maden}, {Meier}, {Morlok}, {Pack}, {Ruf}, {Schmitt-Kopplin},
  {Sch{\"o}Nb{\"a}Chler}, {Steele}, {Spurn{\'y}}, and
  {Wimmer}]{2017M&PS...52.1683B}
A.~{Bischoff}, J.-A. {Barrat}, K.~{Bauer}, C.~{Burkhardt}, H.~{Busemann},
  S.~{Ebert}, M.~{Gonsior}, J.~{Hakenm{\"u}ller}, J.~{Haloda}, D.~{Harries},
  D.~{Heinlein}, H.~{Hiesinger}, R.~{Hochleitner}, V.~{Hoffmann},
  M.~{Kaliwoda}, M.~{Laubenstein}, C.~{Maden}, M.~M.~M. {Meier}, A.~{Morlok},
  A.~{Pack}, A.~{Ruf}, P.~{Schmitt-Kopplin}, M.~{Sch{\"o}Nb{\"a}Chler},
  R.~C.~J. {Steele}, P.~{Spurn{\'y}}, and K.~{Wimmer}.
\newblock {The Stubenberg meteorite -- An LL6 chondrite fragmental breccia
  recovered soon after precise prediction of the strewn field}.
\newblock \emph{Meteoritics and Planetary Science}, 52:\penalty0 1683--1703,
  Aug. 2017.
\newblock \doi{10.1111/maps.12883}.

\bibitem[{Bland} et~al.(2009){Bland}, {Spurn{\'y}}, {Towner}, {Bevan},
  {Singleton}, {Bottke}, {Greenwood}, {Chesley}, {Shrben{\'y}}, {Borovi{\v
  c}ka}, {Ceplecha}, {McClafferty}, {Vaughan}, {Benedix}, {Deacon}, {Howard},
  {Franchi}, and {Hough}]{2009Sci...325.1525B}
P.~A. {Bland}, P.~{Spurn{\'y}}, M.~C. {Towner}, A.~W.~R. {Bevan}, A.~T.
  {Singleton}, W.~F. {Bottke}, R.~C. {Greenwood}, S.~R. {Chesley},
  L.~{Shrben{\'y}}, J.~{Borovi{\v c}ka}, Z.~{Ceplecha}, T.~P. {McClafferty},
  D.~{Vaughan}, G.~K. {Benedix}, G.~{Deacon}, K.~T. {Howard}, I.~A. {Franchi},
  and R.~M. {Hough}.
\newblock {An Anomalous Basaltic Meteorite from the Innermost Main Belt}.
\newblock \emph{Science}, 325:\penalty0 1525, Sept. 2009.
\newblock \doi{10.1126/science.1174787}.

\bibitem[{Bland} et~al.(2012){Bland}, {Spurn{\'y}}, {Bevan}, {Howard},
  {Towner}, {Benedix}, {Greenwood}, {Shrben{\'y}}, {Franchi}, {Deacon},
  {Borovi{\v c}ka}, {Ceplecha}, {Vaughan}, and {Hough}]{2012AuJES..59..177B}
P.~A. {Bland}, P.~{Spurn{\'y}}, A.~W.~R. {Bevan}, K.~T. {Howard}, M.~C.
  {Towner}, G.~K. {Benedix}, R.~C. {Greenwood}, L.~{Shrben{\'y}}, I.~A.
  {Franchi}, G.~{Deacon}, J.~{Borovi{\v c}ka}, Z.~{Ceplecha}, D.~{Vaughan}, and
  R.~M. {Hough}.
\newblock {The Australian Desert Fireball Network: a new era for planetary
  science}.
\newblock \emph{Australian Journal of Earth Sciences}, 59:\penalty0 177--187,
  Mar. 2012.
\newblock \doi{10.1080/08120099.2011.595428}.

\bibitem[{Borovi{\v c}ka}(1990)]{1990BAICz..41..391B}
J.~{Borovi{\v c}ka}.
\newblock {The comparison of two methods of determining meteor trajectories
  from photographs}.
\newblock \emph{Bulletin of the Astronomical Institutes of Czechoslovakia},
  41:\penalty0 391--396, Dec. 1990.

\bibitem[{Borovi{\v c}ka} et~al.(2003){Borovi{\v c}ka}, {Spurn{\'y}},
  {Kalenda}, and {Tagliaferri}]{2003M&PS...38..975B}
J.~{Borovi{\v c}ka}, P.~{Spurn{\'y}}, P.~{Kalenda}, and E.~{Tagliaferri}.
\newblock {The Mor{\'a}vka meteorite fall: 1 Description of the events and
  determination of the fireball trajectory and orbit from video records}.
\newblock \emph{Meteoritics and Planetary Science}, 38:\penalty0 975--987, July
  2003.
\newblock \doi{10.1111/j.1945-5100.2003.tb00293.x}.

\bibitem[{Borovi{\v c}ka} et~al.(2013){Borovi{\v c}ka}, {Spurn{\'y}}, {Brown},
  {Wiegert}, {Kalenda}, {Clark}, and {Shrben{\'y}}]{2013Natur.503..235B}
J.~{Borovi{\v c}ka}, P.~{Spurn{\'y}}, P.~{Brown}, P.~{Wiegert}, P.~{Kalenda},
  D.~{Clark}, and L.~{Shrben{\'y}}.
\newblock {The trajectory, structure and origin of the Chelyabinsk asteroidal
  impactor}.
\newblock \emph{\nat}, 503:\penalty0 235--237, Nov. 2013.
\newblock \doi{10.1038/nature12671}.

\bibitem[{Borovi{\v c}ka} et~al.(2015){Borovi{\v c}ka}, {Spurn{\'y}}, and
  {Brown}]{2015aste.book..257B}
J.~{Borovi{\v c}ka}, P.~{Spurn{\'y}}, and P.~{Brown}.
\newblock \emph{{Small Near-Earth Asteroids as a Source of Meteorites}}, pages
  257--280.
\newblock 2015.
\newblock \doi{10.2458/azu_uapress_9780816532131-ch014}.

\bibitem[{Bowell} et~al.(2002){Bowell}, {Virtanen}, {Muinonen}, and
  {Boattini}]{2002aste.book...27B}
E.~{Bowell}, J.~{Virtanen}, K.~{Muinonen}, and A.~{Boattini}.
\newblock \emph{{Asteroid Orbit Computation}}, pages 27--43.
\newblock 2002.

\bibitem[Bronshten(1983)]{Bronshten1983}
V.~A. Bronshten.
\newblock \emph{{Physics of Meteoric Phenomena}}.
\newblock Geophysics and Astrophysics Monographs. Reidel, Dordrecht,
  Netherlands, 1983.
\newblock ISBN 9789027716545.

\bibitem[{Brown} et~al.(2004){Brown}, {Pack}, {Edwards}, {Revelle}, {Yoo},
  {Spalding}, and {Tagliaferri}]{2004M&PS...39.1781B}
P.~{Brown}, D.~{Pack}, W.~N. {Edwards}, D.~O. {Revelle}, B.~B. {Yoo}, R.~E.
  {Spalding}, and E.~{Tagliaferri}.
\newblock {The orbit, atmospheric dynamics, and initial mass of the Park Forest
  meteorite}.
\newblock \emph{Meteoritics and Planetary Science}, 39:\penalty0 1781--1796,
  Nov. 2004.
\newblock \doi{10.1111/j.1945-5100.2004.tb00075.x}.

\bibitem[{Brown} et~al.(2011){Brown}, {McCausland}, {Fries}, {Silber},
  {Edwards}, {Wong}, {Weryk}, {Fries}, and {Krzeminski}]{2011M&PS...46..339B}
P.~{Brown}, P.~J.~A. {McCausland}, M.~{Fries}, E.~{Silber}, W.~N. {Edwards},
  D.~K. {Wong}, R.~J. {Weryk}, J.~{Fries}, and Z.~{Krzeminski}.
\newblock {The fall of the Grimsby meteorite I: Fireball dynamics and orbit
  from radar, video, and infrasound records}.
\newblock \emph{Meteoritics and Planetary Science}, 46:\penalty0 339--363, Mar.
  2011.
\newblock \doi{10.1111/j.1945-5100.2010.01167.x}.

\bibitem[{Brown} et~al.(2013){Brown}, {Assink}, {Astiz}, {Blaauw}, {Boslough},
  {Borovi{\v c}ka}, {Brachet}, {Brown}, {Campbell-Brown}, {Ceranna}, {Cooke},
  {de Groot-Hedlin}, {Drob}, {Edwards}, {Evers}, {Garces}, {Gill}, {Hedlin},
  {Kingery}, {Laske}, {Le Pichon}, {Mialle}, {Moser}, {Saffer}, {Silber},
  {Smets}, {Spalding}, {Spurn{\'y}}, {Tagliaferri}, {Uren}, {Weryk},
  {Whitaker}, and {Krzeminski}]{2013Natur.503..238B}
P.~G. {Brown}, J.~D. {Assink}, L.~{Astiz}, R.~{Blaauw}, M.~B. {Boslough},
  J.~{Borovi{\v c}ka}, N.~{Brachet}, D.~{Brown}, M.~{Campbell-Brown},
  L.~{Ceranna}, W.~{Cooke}, C.~{de Groot-Hedlin}, D.~P. {Drob}, W.~{Edwards},
  L.~G. {Evers}, M.~{Garces}, J.~{Gill}, M.~{Hedlin}, A.~{Kingery}, G.~{Laske},
  A.~{Le Pichon}, P.~{Mialle}, D.~E. {Moser}, A.~{Saffer}, E.~{Silber},
  P.~{Smets}, R.~E. {Spalding}, P.~{Spurn{\'y}}, E.~{Tagliaferri}, D.~{Uren},
  R.~J. {Weryk}, R.~{Whitaker}, and Z.~{Krzeminski}.
\newblock {A 500-kiloton airburst over Chelyabinsk and an enhanced hazard from
  small impactors}.
\newblock \emph{\nat}, 503:\penalty0 238--241, Nov. 2013.
\newblock \doi{10.1038/nature12741}.

\bibitem[{Ceplecha}(1961)]{1961BAICz..12...21C}
Z.~{Ceplecha}.
\newblock {Multiple fall of P\v{r}\'{\i}bram meteorites photographed. 1.
  Double-station photographs of the fireball and their relations to the found
  meteorites}.
\newblock \emph{Bulletin of the Astronomical Institutes of Czechoslovakia},
  12:\penalty0 21, 1961.

\bibitem[{Devillepoix} et~al.(2016){Devillepoix}, {Bland}, {Towner},
  {Cup{\'a}k}, {Sansom}, {Jansen-Sturgeon}, {Howie}, {Paxman}, and
  {Hartig}]{2016pimo.conf...60D}
H.~A.~R. {Devillepoix}, P.~A. {Bland}, M.~C. {Towner}, M.~{Cup{\'a}k}, E.~K.
  {Sansom}, T.~{Jansen-Sturgeon}, R.~M. {Howie}, J.~{Paxman}, and B.~A.~D.
  {Hartig}.
\newblock {Status of the Desert Fireball Network}.
\newblock In A.~{Roggemans} and P.~{Roggemans}, editors, \emph{International
  Meteor Conference Egmond, the Netherlands, 2-5 June 2016}, pages 60--62, Jan.
  2016.

\bibitem[{Farinella} et~al.(1998){Farinella}, {Vokrouhlick{\'y}}, and
  {Hartmann}]{1998Icar..132..378F}
P.~{Farinella}, D.~{Vokrouhlick{\'y}}, and W.~K. {Hartmann}.
\newblock {Meteorite Delivery via Yarkovsky Orbital Drift}.
\newblock \emph{\icarus}, 132:\penalty0 378--387, Apr. 1998.
\newblock \doi{10.1006/icar.1997.5872}.

\bibitem[{Fries} and {Fries}(2010)]{2010M&PS...45.1476F}
M.~{Fries} and J.~{Fries}.
\newblock {Doppler weather radar as a meteorite recovery tool}.
\newblock \emph{Meteoritics and Planetary Science}, 45:\penalty0 1476--1487,
  Sept. 2010.
\newblock \doi{10.1111/j.1945-5100.2010.01115.x}.

\bibitem[{Fries} et~al.(2014){Fries}, {Le Corre}, {Hankey}, {Fries}, {Matson},
  {Schaefer}, and {Reddy}]{2014M&PS...49.1989F}
M.~{Fries}, L.~{Le Corre}, M.~{Hankey}, J.~{Fries}, R.~{Matson}, J.~{Schaefer},
  and V.~{Reddy}.
\newblock {Detection and rapid recovery of the Sutter's Mill meteorite fall as
  a model for future recoveries worldwide}.
\newblock \emph{Meteoritics and Planetary Science}, 49:\penalty0 1989--1996,
  Nov. 2014.
\newblock \doi{10.1111/maps.12249}.

\bibitem[{Gritsevich}(2009)]{2009AdSpR..44..323G}
M.~I. {Gritsevich}.
\newblock {Determination of parameters of meteor bodies based on flight
  observational data}.
\newblock \emph{Advances in Space Research}, 44:\penalty0 323--334, Aug. 2009.
\newblock \doi{10.1016/j.asr.2009.03.030}.

\bibitem[{Halliday} et~al.(1981){Halliday}, {Griffin}, and
  {Blackwell}]{1981Metic..16..153H}
I.~{Halliday}, A.~A. {Griffin}, and A.~T. {Blackwell}.
\newblock {The Innisfree meteorite fall - A photographic analysis of
  fragmentation, dynamics and luminosity}.
\newblock \emph{Meteoritics}, 16:\penalty0 153--170, June 1981.

\bibitem[{Halliday} et~al.(1996){Halliday}, {Griffin}, and
  {Blackwell}]{1996M&PS...31..185H}
I.~{Halliday}, A.~A. {Griffin}, and A.~T. {Blackwell}.
\newblock {Detailed data for 259 fireballs from the Canadian camera network and
  inferences concerning the influx of large meteoroids}.
\newblock \emph{Meteoritics and Planetary Science}, 31:\penalty0 185--217, Mar.
  1996.
\newblock \doi{10.1111/j.1945-5100.1996.tb02014.x}.

\bibitem[{Hankey} and {Perlerin}(2015)]{2015JIMO...43....2H}
M.~{Hankey} and V.~{Perlerin}.
\newblock {IMO's new online fireball form}.
\newblock \emph{WGN, Journal of the International Meteor Organization},
  43:\penalty0 2--7, Feb. 2015.

\bibitem[Hoppe(1937)]{hoppe1937physikalischen}
J.~Hoppe.
\newblock {Die physikalischen Vorg{\"a}nge beim Eindringen meteoritischer
  K{\"o}rper in die Erdatmosph{\"a}re}.
\newblock \emph{Astronomische Nachrichten}, 262\penalty0 (10):\penalty0
  169--198, 1937.

\bibitem[{Howie} et~al.(2017{\natexlab{a}}){Howie}, {Paxman}, {Bland},
  {Towner}, {Cup\'{a}k}, {Sansom}, and {Devillepoix}]{2017ExA...tmp...19H}
R.~M. {Howie}, J.~{Paxman}, P.~A. {Bland}, M.~C. {Towner}, M.~{Cup\'{a}k},
  E.~K. {Sansom}, and H.~A.~R. {Devillepoix}.
\newblock {How to build a continental scale fireball camera network}.
\newblock \emph{Experimental Astronomy}, May 2017{\natexlab{a}}.
\newblock \doi{10.1007/s10686-017-9532-7}.

\bibitem[{Howie} et~al.(2017{\natexlab{b}}){Howie}, {Paxman}, {Bland},
  {Towner}, {Sansom}, and {Devillepoix}]{2017M&PS...52.1669H}
R.~M. {Howie}, J.~{Paxman}, P.~A. {Bland}, M.~C. {Towner}, E.~K. {Sansom}, and
  H.~A.~R. {Devillepoix}.
\newblock {Submillisecond fireball timing using de Bruijn timecodes}.
\newblock \emph{Meteoritics and Planetary Science}, 52:\penalty0 1669--1682,
  Aug. 2017{\natexlab{b}}.
\newblock \doi{10.1111/maps.12878}.

\bibitem[{Ivezic} et~al.(2008){Ivezic}, {Axelrod}, {Brandt}, {Burke}, {Claver},
  {Connolly}, {Cook}, {Gee}, {Gilmore}, {Jacoby}, {Jones}, {Kahn}, {Kantor},
  {Krabbendam}, {Lupton}, {Monet}, {Pinto}, {Saha}, {Schalk}, {Schneider},
  {Strauss}, {Stubbs}, {Sweeney}, {Szalay}, {Thaler}, {Tyson}, and {LSST
  Collaboration}]{2008SerAJ.176....1I}
Z.~{Ivezic}, T.~{Axelrod}, W.~N. {Brandt}, D.~L. {Burke}, C.~F. {Claver},
  A.~{Connolly}, K.~H. {Cook}, P.~{Gee}, D.~K. {Gilmore}, S.~H. {Jacoby}, R.~L.
  {Jones}, S.~M. {Kahn}, J.~P. {Kantor}, V.~V. {Krabbendam}, R.~H. {Lupton},
  D.~G. {Monet}, P.~A. {Pinto}, A.~{Saha}, T.~L. {Schalk}, D.~P. {Schneider},
  M.~A. {Strauss}, C.~W. {Stubbs}, D.~{Sweeney}, A.~{Szalay}, J.~J. {Thaler},
  J.~A. {Tyson}, and {LSST Collaboration}.
\newblock {Large Synoptic Survey Telescope: From Science Drivers To Reference
  Design}.
\newblock \emph{Serbian Astronomical Journal}, 176:\penalty0 1--13, June 2008.
\newblock \doi{10.2298/SAJ0876001I}.

\bibitem[{Jenniskens}(2014)]{2014me13.conf...57J}
P.~{Jenniskens}.
\newblock {Recent documented meteorite falls, a review of meteorite - asteroid
  links}.
\newblock \emph{Meteoroids 2013}, July 2014.

\bibitem[{Jenniskens} et~al.(2009){Jenniskens}, {Shaddad}, {Numan}, {Elsir},
  {Kudoda}, {Zolensky}, {Le}, {Robinson}, {Friedrich}, {Rumble}, {Steele},
  {Chesley}, {Fitzsimmons}, {Duddy}, {Hsieh}, {Ramsay}, {Brown}, {Edwards},
  {Tagliaferri}, {Boslough}, {Spalding}, {Dantowitz}, {Kozubal}, {Pravec},
  {Borovi{\v c}ka}, {Charvat}, {Vaubaillon}, {Kuiper}, {Albers}, {Bishop},
  {Mancinelli}, {Sandford}, {Milam}, {Nuevo}, and
  {Worden}]{2009Natur.458..485J}
P.~{Jenniskens}, M.~H. {Shaddad}, D.~{Numan}, S.~{Elsir}, A.~M. {Kudoda}, M.~E.
  {Zolensky}, L.~{Le}, G.~A. {Robinson}, J.~M. {Friedrich}, D.~{Rumble},
  A.~{Steele}, S.~R. {Chesley}, A.~{Fitzsimmons}, S.~{Duddy}, H.~H. {Hsieh},
  G.~{Ramsay}, P.~G. {Brown}, W.~N. {Edwards}, E.~{Tagliaferri}, M.~B.
  {Boslough}, R.~E. {Spalding}, R.~{Dantowitz}, M.~{Kozubal}, P.~{Pravec},
  J.~{Borovi{\v c}ka}, Z.~{Charvat}, J.~{Vaubaillon}, J.~{Kuiper}, J.~{Albers},
  J.~L. {Bishop}, R.~L. {Mancinelli}, S.~A. {Sandford}, S.~N. {Milam},
  M.~{Nuevo}, and S.~P. {Worden}.
\newblock {The impact and recovery of asteroid 2008 TC$_{3}$}.
\newblock \emph{\nat}, 458:\penalty0 485--488, Mar. 2009.
\newblock \doi{10.1038/nature07920}.

\bibitem[{Jenniskens} et~al.(2012){Jenniskens}, {Fries}, {Yin}, {Zolensky},
  {Krot}, {Sandford}, {Sears}, {Beauford}, {Ebel}, {Friedrich}, {Nagashima},
  {Wimpenny}, {Yamakawa}, {Nishiizumi}, {Hamajima}, {Caffee}, {Welten},
  {Laubenstein}, {Davis}, {Simon}, {Heck}, {Young}, {Kohl}, {Thiemens}, {Nunn},
  {Mikouchi}, {Hagiya}, {Ohsumi}, {Cahill}, {Lawton}, {Barnes}, {Steele},
  {Rochette}, {Verosub}, {Gattacceca}, {Cooper}, {Glavin}, {Burton}, {Dworkin},
  {Elsila}, {Pizzarello}, {Ogliore}, {Schmitt-Kopplin}, {Harir}, {Hertkorn},
  {Verchovsky}, {Grady}, {Nagao}, {Okazaki}, {Takechi}, {Hiroi}, {Smith},
  {Silber}, {Brown}, {Albers}, {Klotz}, {Hankey}, {Matson}, {Fries}, {Walker},
  {Puchtel}, {Lee}, {Erdman}, {Eppich}, {Roeske}, {Gabelica}, {Lerche},
  {Nuevo}, {Girten}, and {Worden}]{2012Sci...338.1583J}
P.~{Jenniskens}, M.~D. {Fries}, Q.-Z. {Yin}, M.~{Zolensky}, A.~N. {Krot}, S.~A.
  {Sandford}, D.~{Sears}, R.~{Beauford}, D.~S. {Ebel}, J.~M. {Friedrich},
  K.~{Nagashima}, J.~{Wimpenny}, A.~{Yamakawa}, K.~{Nishiizumi}, Y.~{Hamajima},
  M.~W. {Caffee}, K.~C. {Welten}, M.~{Laubenstein}, A.~M. {Davis}, S.~B.
  {Simon}, P.~R. {Heck}, E.~D. {Young}, I.~E. {Kohl}, M.~H. {Thiemens}, M.~H.
  {Nunn}, T.~{Mikouchi}, K.~{Hagiya}, K.~{Ohsumi}, T.~A. {Cahill}, J.~A.
  {Lawton}, D.~{Barnes}, A.~{Steele}, P.~{Rochette}, K.~L. {Verosub},
  J.~{Gattacceca}, G.~{Cooper}, D.~P. {Glavin}, A.~S. {Burton}, J.~P.
  {Dworkin}, J.~E. {Elsila}, S.~{Pizzarello}, R.~{Ogliore},
  P.~{Schmitt-Kopplin}, M.~{Harir}, N.~{Hertkorn}, A.~{Verchovsky}, M.~{Grady},
  K.~{Nagao}, R.~{Okazaki}, H.~{Takechi}, T.~{Hiroi}, K.~{Smith}, E.~A.
  {Silber}, P.~G. {Brown}, J.~{Albers}, D.~{Klotz}, M.~{Hankey}, R.~{Matson},
  J.~A. {Fries}, R.~J. {Walker}, I.~{Puchtel}, C.-T.~A. {Lee}, M.~E. {Erdman},
  G.~R. {Eppich}, S.~{Roeske}, Z.~{Gabelica}, M.~{Lerche}, M.~{Nuevo},
  B.~{Girten}, and S.~P. {Worden}.
\newblock {Radar-Enabled Recovery of the Sutter's Mill Meteorite, a
  Carbonaceous Chondrite Regolith Breccia}.
\newblock \emph{Science}, 338:\penalty0 1583, Dec. 2012.
\newblock \doi{10.1126/science.1227163}.

\bibitem[{Jenniskens} et~al.(2014){Jenniskens}, {Rubin}, {Yin}, {Sears},
  {Sandford}, {Zolensky}, {Krot}, {Blair}, {Kane}, {Utas}, {Verish},
  {Friedrich}, {Wimpenny}, {Eppich}, {Ziegler}, {Verosub}, {Rowland}, {Albers},
  {Gural}, {Grigsby}, {Fries}, {Matson}, {Johnston}, {Silber}, {Brown},
  {Yamakawa}, {Sanborn}, {Laubenstein}, {Welten}, {Nishiizumi}, {Meier},
  {Busemann}, {Clay}, {Caffee}, {Schmitt-Kopplin}, {Hertkorn}, {Glavin},
  {Callahan}, {Dworkin}, {Wu}, {Zare}, {Grady}, {Verchovsky}, {Emel'Yanenko},
  {Naroenkov}, {Clark}, {Girten}, and {Worden}]{2014M&PS...49.1388J}
P.~{Jenniskens}, A.~E. {Rubin}, Q.-Z. {Yin}, D.~W.~G. {Sears}, S.~A.
  {Sandford}, M.~E. {Zolensky}, A.~N. {Krot}, L.~{Blair}, D.~{Kane}, J.~{Utas},
  R.~{Verish}, J.~M. {Friedrich}, J.~{Wimpenny}, G.~R. {Eppich}, K.~{Ziegler},
  K.~L. {Verosub}, D.~J. {Rowland}, J.~{Albers}, P.~S. {Gural}, B.~{Grigsby},
  M.~D. {Fries}, R.~{Matson}, M.~{Johnston}, E.~{Silber}, P.~{Brown},
  A.~{Yamakawa}, M.~E. {Sanborn}, M.~{Laubenstein}, K.~C. {Welten},
  K.~{Nishiizumi}, M.~M.~M. {Meier}, H.~{Busemann}, P.~{Clay}, M.~W. {Caffee},
  P.~{Schmitt-Kopplin}, N.~{Hertkorn}, D.~P. {Glavin}, M.~P. {Callahan}, J.~P.
  {Dworkin}, Q.~{Wu}, R.~N. {Zare}, M.~{Grady}, S.~{Verchovsky},
  V.~{Emel'Yanenko}, S.~{Naroenkov}, D.~L. {Clark}, B.~{Girten}, and P.~S.
  {Worden}.
\newblock {Fall, recovery, and characterization of the Novato L6 chondrite
  breccia}.
\newblock \emph{Meteoritics and Planetary Science}, 49:\penalty0 1388--1425,
  Aug. 2014.
\newblock \doi{10.1111/maps.12323}.

\bibitem[Keay(1992)]{keay1992electrophonic}
C.~S. Keay.
\newblock Electrophonic sounds from large meteor fireballs.
\newblock \emph{Meteoritics \& Planetary Science}, 27\penalty0 (2):\penalty0
  144--148, 1992.

\bibitem[{Keil} et~al.(1997){Keil}, {Stoeffler}, {Love}, and
  {Scott}]{1997M&PS...32..349K}
K.~{Keil}, D.~{Stoeffler}, S.~G. {Love}, and E.~R.~D. {Scott}.
\newblock {Constraints on the role of impact heating and melting in asteroids}.
\newblock \emph{Meteoritics and Planetary Science}, 32, May 1997.
\newblock \doi{10.1111/j.1945-5100.1997.tb01278.x}.

\bibitem[McCrosky and Boeschenstein(1965)]{McCrosky1965}
R.~E. McCrosky and H.~Boeschenstein.
\newblock {The Prairie Meteorite Network}.
\newblock \emph{Optical Engineering}, 3\penalty0 (4):\penalty0 304127--304127,
  1965.

\bibitem[{Morbidelli} et~al.(1994){Morbidelli}, {Gonczi}, {Froeschle}, and
  {Farinella}]{1994A&A...282..955M}
A.~{Morbidelli}, R.~{Gonczi}, C.~{Froeschle}, and P.~{Farinella}.
\newblock {Delivery of meteorites through the nu$_{6}$ secular resonance}.
\newblock \emph{\aap}, 282:\penalty0 955--979, Feb. 1994.

\bibitem[{Nakamura} et~al.(2011){Nakamura}, {Noguchi}, {Tanaka}, {Zolensky},
  {Kimura}, {Tsuchiyama}, {Nakato}, {Ogami}, {Ishida}, {Uesugi}, {Yada},
  {Shirai}, {Fujimura}, {Okazaki}, {Sandford}, {Ishibashi}, {Abe}, {Okada},
  {Ueno}, {Mukai}, {Yoshikawa}, and {Kawaguchi}]{2011Sci...333.1113N}
T.~{Nakamura}, T.~{Noguchi}, M.~{Tanaka}, M.~E. {Zolensky}, M.~{Kimura},
  A.~{Tsuchiyama}, A.~{Nakato}, T.~{Ogami}, H.~{Ishida}, M.~{Uesugi},
  T.~{Yada}, K.~{Shirai}, A.~{Fujimura}, R.~{Okazaki}, S.~A. {Sandford},
  Y.~{Ishibashi}, M.~{Abe}, T.~{Okada}, M.~{Ueno}, T.~{Mukai}, M.~{Yoshikawa},
  and J.~{Kawaguchi}.
\newblock {Itokawa Dust Particles: A Direct Link Between S-Type Asteroids and
  Ordinary Chondrites}.
\newblock \emph{Science}, 333:\penalty0 1113, Aug. 2011.
\newblock \doi{10.1126/science.1207758}.

\bibitem[{Nesvorn{\'y}} et~al.(2009){Nesvorn{\'y}}, {Vokrouhlick{\'y}},
  {Morbidelli}, and {Bottke}]{2009Icar..200..698N}
D.~{Nesvorn{\'y}}, D.~{Vokrouhlick{\'y}}, A.~{Morbidelli}, and W.~F. {Bottke}.
\newblock {Asteroidal source of L chondrite meteorites}.
\newblock \emph{\icarus}, 200:\penalty0 698--701, Apr. 2009.
\newblock \doi{10.1016/j.icarus.2008.12.016}.

\bibitem[{Oberst} et~al.(1998){Oberst}, {Molau}, {Heinlein}, {Gritzner},
  {Schindler}, {Spurny}, {Ceplecha}, {Rendtel}, and
  {Betlem}]{1998M&PS...33...49O}
J.~{Oberst}, S.~{Molau}, D.~{Heinlein}, C.~{Gritzner}, M.~{Schindler},
  P.~{Spurny}, Z.~{Ceplecha}, J.~{Rendtel}, and H.~{Betlem}.
\newblock {The ``European Fireball Network'': Current status and future
  prospects}.
\newblock \emph{Meteoritics and Planetary Science}, 33, Jan. 1998.
\newblock \doi{10.1111/j.1945-5100.1998.tb01606.x}.

\bibitem[{Palotai} et~al.(2018){Palotai}, {Sankar}, {Free}, {Howell},
  {Botella}, and {Batcheldor}]{2018arXiv180105072P}
C.~{Palotai}, R.~{Sankar}, D.~L. {Free}, J.~A. {Howell}, E.~{Botella}, and
  D.~{Batcheldor}.
\newblock {Analysis of June 2, 2016 bolide event}.
\newblock \emph{ArXiv e-prints}, Jan. 2018.

\bibitem[Paxman and Bland(2014)]{2014LPI....45.1731P}
J.~Paxman and P.~Bland.
\newblock Fireballs in the sky: Improving the accuracy of crowd sourced
  fireball observation through the application of smartphone technology.
\newblock In \emph{Lunar and Planetary Science Conference}, volume~45, page
  1731, 2014.

\bibitem[{Picone} et~al.(2002){Picone}, {Hedin}, {Drob}, and
  {Aikin}]{2002JGRA..107.1468P}
J.~M. {Picone}, A.~E. {Hedin}, D.~P. {Drob}, and A.~C. {Aikin}.
\newblock {NRLMSISE-00 empirical model of the atmosphere: Statistical
  comparisons and scientific issues}.
\newblock \emph{Journal of Geophysical Research (Space Physics)}, 107:\penalty0
  1468, Dec. 2002.
\newblock \doi{10.1029/2002JA009430}.

\bibitem[{Reddy} et~al.(2014){Reddy}, {Sanchez}, {Bottke}, {Cloutis}, {Izawa},
  {O'Brien}, {Mann}, {Cuddy}, {Le Corre}, {Gaffey}, and
  {Fujihara}]{2014Icar..237..116R}
V.~{Reddy}, J.~A. {Sanchez}, W.~F. {Bottke}, E.~A. {Cloutis}, M.~R.~M. {Izawa},
  D.~P. {O'Brien}, P.~{Mann}, M.~{Cuddy}, L.~{Le Corre}, M.~J. {Gaffey}, and
  G.~{Fujihara}.
\newblock {Chelyabinsk meteorite explains unusual spectral properties of
  Baptistina Asteroid Family}.
\newblock \emph{\icarus}, 237:\penalty0 116--130, July 2014.
\newblock \doi{10.1016/j.icarus.2014.04.027}.

\bibitem[{Rein} and {Tamayo}(2015)]{2015MNRAS.452..376R}
H.~{Rein} and D.~{Tamayo}.
\newblock {WHFAST: a fast and unbiased implementation of a symplectic
  Wisdom-Holman integrator for long-term gravitational simulations}.
\newblock \emph{\mnras}, 452:\penalty0 376--388, Sept. 2015.
\newblock \doi{10.1093/mnras/stv1257}.

\bibitem[{Sansom} et~al.(2016){Sansom}, {Ridgewell}, {Bland}, and
  {Paxman}]{2016pimo.conf..267S}
E.~{Sansom}, J.~{Ridgewell}, P.~{Bland}, and J.~{Paxman}.
\newblock {Meteor reporting made easy- The Fireballs in the Sky smartphone
  app}.
\newblock In A.~{Roggemans} and P.~{Roggemans}, editors, \emph{International
  Meteor Conference Egmond, the Netherlands, 2-5 June 2016}, pages 267--269,
  Jan. 2016.

\bibitem[Sansom(2016)]{sansom2016tracking}
E.~K. Sansom.
\newblock \emph{Tracking Meteoroids in the Atmosphere: Fireball Trajectory
  Analysis}.
\newblock PhD thesis, Curtin University, 2016.
\newblock URL \url{+ http://hdl.handle.net/20.500.11937/55061}.

\bibitem[{Sansom} et~al.(2015){Sansom}, {Bland}, {Paxman}, and
  {Towner}]{2015M&PS...50.1423S}
E.~K. {Sansom}, P.~{Bland}, J.~{Paxman}, and M.~{Towner}.
\newblock {A novel approach to fireball modeling: The observable and the
  calculated}.
\newblock \emph{Meteoritics and Planetary Science}, 50:\penalty0 1423--1435,
  Aug. 2015.
\newblock \doi{10.1111/maps.12478}.

\bibitem[{Sansom} et~al.(2017){Sansom}, {Rutten}, and
  {Bland}]{2017AJ....153...87S}
E.~K. {Sansom}, M.~G. {Rutten}, and P.~A. {Bland}.
\newblock {Analyzing Meteoroid Flights Using Particle Filters}.
\newblock \emph{\aj}, 153:\penalty0 87, Feb. 2017.
\newblock \doi{10.3847/1538-3881/153/2/87}.

\bibitem[{Schmitz} et~al.(2001){Schmitz}, {Tassinari}, and
  {Peucker-Ehrenbrink}]{2001E&PSL.194....1S}
B.~{Schmitz}, M.~{Tassinari}, and B.~{Peucker-Ehrenbrink}.
\newblock {A rain of ordinary chondritic meteorites in the early Ordovician}.
\newblock \emph{Earth and Planetary Science Letters}, 194:\penalty0 1--15, Dec.
  2001.
\newblock \doi{10.1016/S0012-821X(01)00559-3}.

\bibitem[{Skamarock} et~al.(2008){Skamarock}, {Klemp}, {Dudhia}, {Gill},
  {Barker}, {Duda}, {Huang}, {Wang}, and {Powers}]{skamarock2008description}
W.~C. {Skamarock}, J.~B. {Klemp}, J.~{Dudhia}, D.~O. {Gill}, D.~M. {Barker},
  M.~G. {Duda}, X.~Y. {Huang}, W.~{Wang}, and J.~G. {Powers}.
\newblock A description of the advanced research wrf version 3.
\newblock Technical report, NCAR Technical Note NCAR/TN-475+STR, 2008.

\bibitem[{Southworth} and {Hawkins}(1963)]{1963SCoA....7..261S}
R.~B. {Southworth} and G.~S. {Hawkins}.
\newblock {Statistics of meteor streams}.
\newblock \emph{Smithsonian Contributions to Astrophysics}, 7:\penalty0 261,
  1963.

\bibitem[{Spurn{\'y}} et~al.(2010){Spurn{\'y}}, {Borovi{\v c}ka}, {Kac},
  {Kalenda}, {Atanackov}, {Kladnik}, {Heinlein}, and
  {Grau}]{2010M&PS...45.1392S}
P.~{Spurn{\'y}}, J.~{Borovi{\v c}ka}, J.~{Kac}, P.~{Kalenda}, J.~{Atanackov},
  G.~{Kladnik}, D.~{Heinlein}, and T.~{Grau}.
\newblock {Analysis of instrumental observations of the Jesenice meteorite fall
  on April 9, 2009}.
\newblock \emph{Meteoritics and Planetary Science}, 45:\penalty0 1392--1407,
  Aug. 2010.
\newblock \doi{10.1111/j.1945-5100.2010.01121.x}.

\bibitem[{Spurn{\'y}} et~al.(2011){Spurn{\'y}}, {Bland}, {Shrben{\'y}},
  {Towner}, {Borovi{\v c}ka}, {Bevan}, and {Vaughan}]{2011M&PSA..74.5101S}
P.~{Spurn{\'y}}, P.~A. {Bland}, L.~{Shrben{\'y}}, M.~C. {Towner}, J.~{Borovi{\v
  c}ka}, A.~W.~R. {Bevan}, and D.~{Vaughan}.
\newblock {The Mason Gully Meteorite Fall in SW Australia: Fireball Trajectory
  and Orbit from Photographic Records}.
\newblock \emph{Meteoritics and Planetary Science Supplement}, 74:\penalty0
  5101, Sept. 2011.

\bibitem[{Spurn{\'y}} et~al.(2014){Spurn{\'y}}, {Haloda}, {Borovi{\v c}ka},
  {Shrben{\'y}}, and {Halodov{\'a}}]{2014A&A...570A..39S}
P.~{Spurn{\'y}}, J.~{Haloda}, J.~{Borovi{\v c}ka}, L.~{Shrben{\'y}}, and
  P.~{Halodov{\'a}}.
\newblock {Reanalysis of the Bene{\v s}ov bolide and recovery of polymict
  breccia meteorites - old mystery solved after 20 years}.
\newblock \emph{\aap}, 570:\penalty0 A39, Oct. 2014.
\newblock \doi{10.1051/0004-6361/201424308}.

\bibitem[{Spurn{\'y}} et~al.(2016){Spurn{\'y}}, {Borovi{\v c}ka}, {Haloda},
  {Shrben{\'y}}, and {Heinlein}]{2016LPICo1921.6221S}
P.~{Spurn{\'y}}, J.~{Borovi{\v c}ka}, J.~{Haloda}, L.~{Shrben{\'y}}, and
  D.~{Heinlein}.
\newblock {Two Very Precisely Instrumentally Documented Meteorite Falls:
  \v{Z}\v{d}\'{a}r nad S\'{a}zavou and Stubenberg - Prediction and Reality}.
\newblock In \emph{79th Annual Meeting of the Meteoritical Society}, volume
  1921 of \emph{LPI Contributions}, page 6221, Aug. 2016.

\bibitem[{Trigo-Rodr{\'{\i}}guez} et~al.(2006){Trigo-Rodr{\'{\i}}guez},
  {Borovi{\v c}ka}, {Spurn{\'y}}, {Ortiz}, {Docobo}, {Castro-Tirado}, and
  {Llorca}]{2006M&PS...41..505T}
J.~M. {Trigo-Rodr{\'{\i}}guez}, J.~{Borovi{\v c}ka}, P.~{Spurn{\'y}}, J.~L.
  {Ortiz}, J.~A. {Docobo}, A.~J. {Castro-Tirado}, and J.~{Llorca}.
\newblock {The Villalbeto de la Pe{\~n}a meteorite fall: II. Determination of
  atmospheric trajectory and orbit}.
\newblock \emph{Meteoritics and Planetary Science}, 41:\penalty0 505--517, Apr.
  2006.
\newblock \doi{10.1111/j.1945-5100.2006.tb00478.x}.

\bibitem[{Trigo-Rodr{\'{\i}}guez} et~al.(2015){Trigo-Rodr{\'{\i}}guez},
  {Lyytinen}, {Gritsevich}, {Moreno-Ib{\'a}{\~n}ez}, {Bottke}, {Williams},
  {Lupovka}, {Dmitriev}, {Kohout}, and {Grokhovsky}]{2015MNRAS.449.2119T}
J.~M. {Trigo-Rodr{\'{\i}}guez}, E.~{Lyytinen}, M.~{Gritsevich},
  M.~{Moreno-Ib{\'a}{\~n}ez}, W.~F. {Bottke}, I.~{Williams}, V.~{Lupovka},
  V.~{Dmitriev}, T.~{Kohout}, and V.~{Grokhovsky}.
\newblock {Orbit and dynamic origin of the recently recovered Annama's H5
  chondrite}.
\newblock \emph{\mnras}, 449:\penalty0 2119--2127, May 2015.
\newblock \doi{10.1093/mnras/stv378}.

\bibitem[{Vernazza} et~al.(2008){Vernazza}, {Binzel}, {Thomas}, {DeMeo}, {Bus},
  {Rivkin}, and {Tokunaga}]{2008Natur.454..858V}
P.~{Vernazza}, R.~P. {Binzel}, C.~A. {Thomas}, F.~E. {DeMeo}, S.~J. {Bus},
  A.~S. {Rivkin}, and A.~T. {Tokunaga}.
\newblock {Compositional differences between meteorites and near-Earth
  asteroids}.
\newblock \emph{\nat}, 454:\penalty0 858--860, Aug. 2008.
\newblock \doi{10.1038/nature07154}.

\bibitem[{Whipple}(1939)]{1939PAAS....9R.136W}
F.~L. {Whipple}.
\newblock {Photographic meteor studies I}.
\newblock In \emph{Publications of the American Astronomical Society}, volume~9
  of \emph{Publications of the American Astronomical Society}, page 136, 1939.

\end{thebibliography}
\bibliographystyle{abbrvnat}

\end{document}